\documentclass[
twocolumn,
aps,
prb,
reprint,
superscriptaddress,
amsmath,amssymb,
nofootinbib,
longbibliography
]{revtex4-2}
\renewcommand{\selectlanguage}[1]{}
\usepackage[pdftex]{graphicx, color}
\usepackage[hidelinks]{hyperref}
\hypersetup{
 setpagesize=false,
 bookmarksnumbered=true,%
 bookmarksopen=true,%
 colorlinks=true,%
 linkcolor=blue,
 citecolor=blue,
}

\graphicspath{{./figure/}{./fig_phase_diagram_t20/}{./fig_phase_diagram_t202/}}

\usepackage{braket}
\usepackage{times,multirow,amsfonts,bm,xspace,pifont}
\usepackage{soul}
\usepackage[normalem]{ulem}

\begin{document}
\newcommand{\rr}{{\bm r}}
\newcommand{\q}{{\bm q}}
\renewcommand{\k}{{\bm k}}
\newcommand*\wien    {\textsc{wien}2k\xspace}
\newcommand*\textred[1]{\textcolor{red}{#1}}
\newcommand*\textblue[1]{\textcolor{blue}{#1}}
\newcommand{\ki}[1]{{\color{red}\st{#1}}}
\newcommand{\sgn}{\mathrm{sgn}\,}
\newcommand{\tr}{\mathrm{tr}\,}
\newcommand{\Tr}{\mathrm{Tr}\,}
\newcommand{\GL}{{\mathrm{GL}}}
\newcommand{\talpha}{{\tilde{\alpha}}}
\newcommand{\tbeta}{{\tilde{\beta}}}
\newcommand{\mathN}{{\mathcal{N}}}
\newcommand{\mathQ}{{\mathcal{Q}}}
\newcommand{\bv}{{\bar{v}}}
\newcommand{\bj}{{\bar{j}}}
\newcommand{\B}{\mathrm{b}}
\newcommand{\BdG}{\mathrm{BdG}}
\newcommand{\diag}{\mathrm{diag}}

\newcommand{\YY}[1]{\textcolor{magenta}{#1}}
\newcommand{\AD}[1]{\textcolor{blue}{#1}}
\newcommand*{\ADS}[1]{\textcolor{blue}{\sout{#1}}}
\newcommand*\YYS[1]{\textcolor{magenta}{\sout{#1}}}
\newcommand{\reply}[1]{\textcolor{red}{#1}}
\newcommand{\replyS}[1]{\textcolor{red}{\sout{#1}}}
\newcommand*\TK[1]{\textcolor{cyan}{#1}}

\newcommand{\QM}{{\mathcal{G}}}

\newcommand{\one}{{(1)\,}}
\newcommand{\two}{{(2)\,}}
\newcommand{\inter}{{{\rm inter}}}
\newcommand{\intra}{{{\rm intra}}}
\newcommand{\conv}{{\mathrm{c}\,}}
\newcommand{\geom}{{\mathrm{g}\,}}
\newcommand{\para}{{\mathrm{para}\,}}
\newcommand{\dia}{{\mathrm{dia}\,}}
\newcommand{\BZ}{{\mathrm{BZ}\,}}
\newcommand{\FS}{{\mathrm{FS}\,}}

\title{Quantum geometry encoded to pair potentials}

\author{Akito Daido}
\affiliation{Department of Physics, Graduate School of Science, Kyoto University, Kyoto 606-8502, Japan}
\email[]{daido@scphys.kyoto-u.ac.jp}
\author{Taisei Kitamura}
\author{Youichi Yanase}
\affiliation{Department of Physics, Graduate School of Science, Kyoto University, Kyoto 606-8502, Japan}
\date{\today}

\begin{abstract}
Bloch wave functions of electrons have properties called quantum geometry, which has recently attracted much attention as the origin of intriguing physical phenomena.
In this paper, we introduce the notion of the quantum-geometric pair potentials (QGPP) based on the generalized band representation and thereby clarify how the quantum geometry of electrons is transferred to the Cooper pairs they form.
QGPP quantifies the deviation of multiband superconductors from an assembly of single-band superconductors and has a direct connection to the quantum-geometric corrections to thermodynamic coefficients.
We also discuss 
their potential ability to
emulate exotic pair potentials and engineer intriguing superconducting phenomena including topological superconductivity.
\end{abstract}

\maketitle
\section{Introduction}
\begin{figure}
    \centering
    \includegraphics[width=0.43\textwidth]{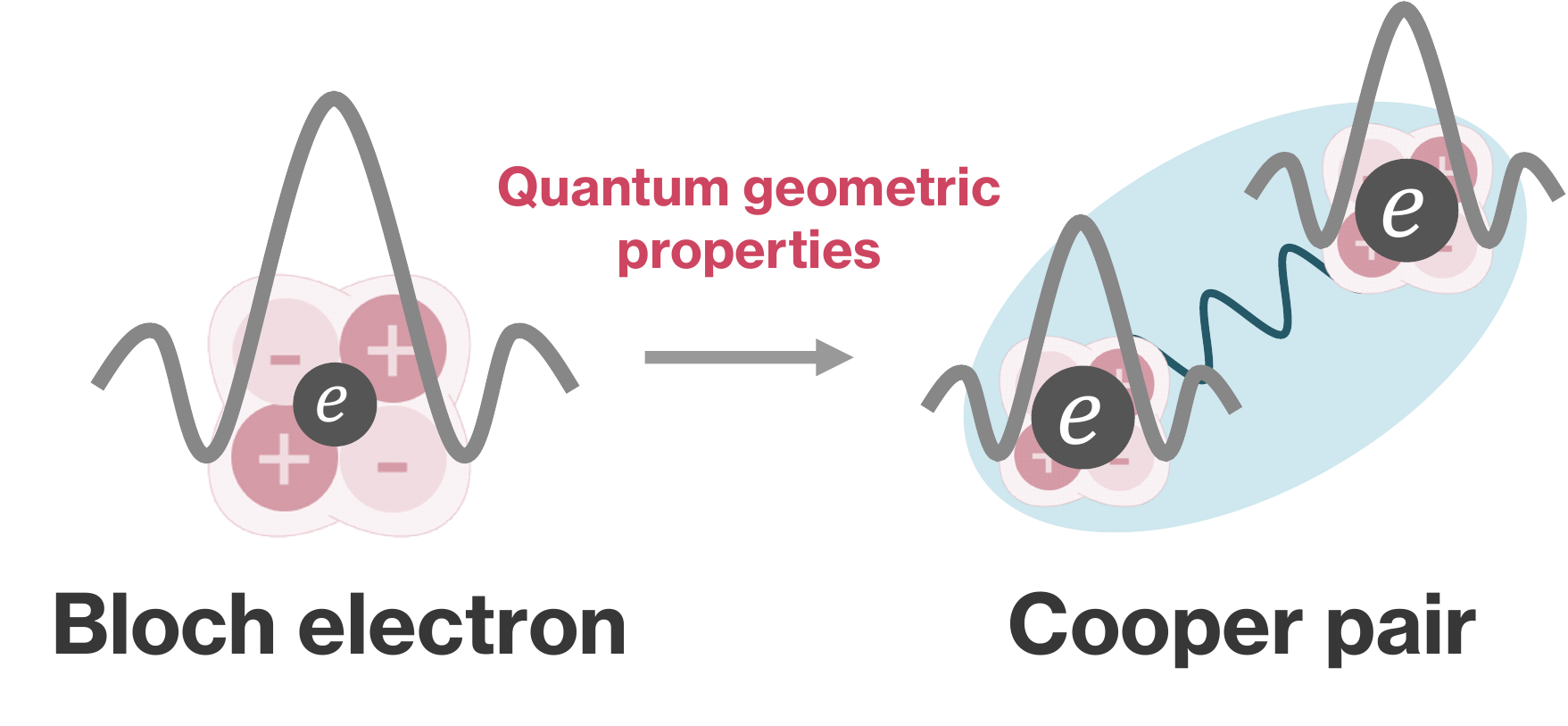}
    \caption{Schematic figure for the quantum geometric properties encoded to Cooper pairs.
    {The configuration of $\pm$ signs in the figure represents the electric quadrupole moment of the Bloch wave packet, i.e., the quantum metric~\cite{Gao2019-hk,Lapa2019-ga}.}
    }
    \label{fig:schematic}
\end{figure}

Recent years have witnessed a variety of exotic superconducting phenomena beyond the Bardeen-Cooper-Schrieffer (BCS) paradigm.
Topological superconductivity (TSC)~\cite{Tanaka2012-vn,Sato2016-sc,Sato2017-lk} is an example, which is characterized by the nontrivial topology of the wave functions of Bogoliubov quasiparticles.
At the early stage of the research, topology coming from the exotic Cooper-pair wave function in unconventional superconductors has mainly been investigated.
On the other hand, it has later been recognized that topology can also originate from the nontrivial Bloch wave function of the normal-state electrons.
For example, Rashba superconductors become TSC under strong Zeeman fields even with $s$-wave pairing, where 
the normal-state Berry curvature is faithfully encoded to the Bogoliubov quasiparticles [Fig.~\ref{fig:schematic}], thereby playing a role similar to the chiral $p$-wave order parameter~\cite{Sato2009-cn,Sato2010-qj}. 
This idea has significantly expanded the research field of TSC, as unconventional superconductors are rare in nature.

Generally speaking, the 
{interesting} properties of the wave functions are quantified by the concept of quantum geometry~\cite{Xiao2010-ah,Resta2011-ob}.
Quantum geometry refers to the nontrivial wave-number (or generally parameter)
dependence of wave functions around each {wave number} such as the Berry curvature, while its global structure in the Brillouin zone gives rise to the topology such as the Chern number.
In superconductors, wave functions of both Bloch electrons and Cooper pairs contribute to quantum geometry, which describes as a whole the quantum geometry of the superconducting states [Fig.~\ref{fig:schematic}].

Quantum geometry is important not only because it gives rise to topology but also because it directly appears in physical phenomena.
Indeed, Berry curvature is known to cause various 
Hall responses in normal and superconducting states~\cite{Xiao2010-ah,Sato2016-sc}, and its multipoles play important roles in nonlinear Hall responses~\cite{Sodemann2015-vj,Ma2019-om,Kang2019-dr,Du2021-fg,Zhang2023-ot,Sankar2024-vx}.
It has also been revealed that the quantum metric, another quantum-geometric quantity, strongly enhances the two-dimensional superconducting transition temperature in twisted bilayer graphene~\cite{Hu2019-uk,Xie2020-iv,Julku2020-wy,Tian2023-ss} and FeSe~\cite{Kitamura2022-pa}.
This follows from the correction to the superfluid weight by quantum geometry, which is overlooked in the standard Fermi-liquid formula~\cite{Peotta2015-tv,Liang2017-ax,Rossi2021-hk,Huhtinen2022-mp,Torma2022-gd}.
The effect of the quantum metric on the coherence length~\cite{Hu2023-vd,Chen2023-km,Iskin2023-cj}, {collective modes~\cite{Villegas2021-gz,Villegas2023-gy},}
finite-momentum superconductivity~\cite{Kitamura2023-on,Kitamura2022-zz,Jiang2023-ga,Chen2023-cg}, {Cooper-pair wave functions~\cite{Herzog-Arbeitman2022-vp},} and pairing interactions~\cite{Kitamura2024-dl,Yu2024-ck} have also been investigated.

The discovery of the quantum-geometric corrections to superconducting properties raises a question about to what extent the textbook formulas of the BCS theory remain valid in existing superconductors.
In particular, a number of non-BCS superconductors have recently been reported even aside from unconventional superconductivity, including superconductors 
in the BCS-Bose-Einstein-condensation (BEC) regime~\cite{
Shibauchi2020-gj,Kasahara2015-sn,Nakagawa2018-cw,Nakagawa2021-ka,Cao2018-hg,Park2021-kk,Kim2022-kw,Chen2024-do}, to which the twisted bilayer graphene belongs.
Quantum geometry is generally developed in systems with multiple degrees of freedom such as orbitals and sublattices.
A comprehensive understanding of the quantum-geometric effects is indispensable to establishing a modern theory of superconductivity.

In this paper, we provide a unified viewpoint on how the quantum-geometric properties of normal-state electrons are encoded to Cooper pairs [Fig.~\ref{fig:schematic}], with particular emphasis on how they affect equilibrium properties.
We clarify the similarities and differences between the single-band model and multi-band systems, the former and the latter of which lack and accompany quantum geometry, respectively.
By introducing the concept of the quantum-geometric pair potential (QGPP), we
{show that QGPP can naturally explain} the quantum-geometric contribution to the thermodynamic properties 
{and} can also be used as a guiding principle to explore{, e.g.,} TSC candidates.
{
QGGP is applied to general systems beyond flat-band ones as well as to general superconducting order parameters. For this reason, we do not assume flat-band dispersion throughout the paper unless otherwise specified.
The general formulation of QGPP
is discussed after initially illustrating QGPP for the plain $s$-wave state (assuming the so-called uniform pairing) in the first few sections.
The obtained formulas are illustrated in a spinful bilayer model for various order parameters with nontrivial matrix structures in the spin and/or layer spaces, whereby we show that QGPP induced by external fields can be used to emulate exotic superconducting states.
}

\section{QGPP in systems of non-degenerate bands}
\label{sec:Abelian}
To clearly illustrate the basic concepts, we first assume non-degenerate energy bands
and the plain $s$-wave superconductivity whose pair potential is of the form
\begin{align}
    \Delta_0\sum_l(c^\dagger_{\bm{k}\uparrow,l}c^\dagger_{-\bm{k}\downarrow,l}-c^\dagger_{\bm{k}\downarrow,l}c^\dagger_{-\bm{k}\uparrow,l}).\label{eq:singlet}
\end{align}
The index $l$ runs over the internal degrees of freedom except for spin such as sublattices and orbitals.
We next discuss the generalization to non-$s$-wave states {in Secs.~\ref{sec:non-s-wave} and~\ref{sec:secIII}, where the assumption of Eq.~\eqref{eq:singlet} is removed and general matrix pair potentials are considered}.

Throughout the paper, we neglect the external-field dependence of the order parameter $\Delta_0$ in the orbital basis.
Such effects{, if necessary,} can be taken into account in our theory by making an additional expansion of the order parameter in the orbital basis in terms of the external field as determined from the gap equation.
This should be independently discussed for each microscopic model as done in Ref.~\onlinecite{Huhtinen2022-mp} since it requires the details of the pairing interaction.

\subsection{Band representation}
Let us introduce Bogoliubov de-Gennes (BdG) Hamiltonian and its band representation.
We consider the mean-field Hamiltonian of the superconducting state, 
\begin{align}
\hat{H}=\frac{1}{2}\sum_{\bm{k}}\Psi^\dagger(\bm{k})H_{\mathrm{BdG}}(\bm{k})\Psi(\bm{k})+\text{const.},
\end{align}
with the Nambu spinor
\begin{align}[\Psi^\dagger(\bm{k})]_l=({c}^\dagger_{\uparrow,l}(\bm{k}),{c}^\dagger_{\downarrow,l}(\bm{k}),{c}_{\downarrow,l}(-\bm{k}),-{c}_{\uparrow,l}(-\bm{k})).
\end{align}
{When we adopt Eq.~\eqref{eq:singlet},} the BdG Hamiltonian in this basis is given by
\begin{align}
    H_{\mathrm{BdG}}(\bm{k})&=\begin{pmatrix}
    H_{\rm N}(\bm{k})&\Delta_0\\
    \Delta_0&-U_\Theta H_{\rm N}(-\bm{k})^TU_\Theta^\dagger
    \end{pmatrix},\label{eq:X_b}
\end{align}
where $H_{\rm N}(\bm{k})$ is the Bloch Hamiltonian of the normal state.
The gap function is proportional to the identity matrix, while the 
spin-singlet wave function $U_{\Theta}=is_y$ in the usual basis is removed by the choice of the Nambu spinor.
Here, $s_i$ represents Pauli matrices in the spin space.
An advantage of this basis is that the hole sector becomes the time-reversal partner of the electron sector: 
$-U_\Theta H_{\rm N}(-\bm{k})^TU_\Theta^\dagger=-\Theta H_{\rm N}(-\bm{k})\Theta^{-1}$,
where $\Theta {\equiv} U_\Theta K$ and $K$ represent the time-reversal and complex-conjugate operators{, and $H_{\rm N}(-\bm{k})^T=H_{\rm N}(-\bm{k})^*$ is used}.
These properties of this basis allow easier access to the analogy with single-band superconductivity.

The band representation is a frequently used approach to understand the properties of the BdG Hamiltonian $H_{\mathrm{BdG}}(\bm{k})$. We start from the normal-state Hamiltonian, which is diagonalized as
\begin{align}
U_{\rm N}(\bm{k})H_{\rm N}(\bm{k})U_{\rm N}^\dagger(\bm{k})\equiv\epsilon(\bm{k}),\  [\epsilon(\bm{k})]_{nm}=\delta_{nm}\epsilon_n(\bm{k}).
\end{align}
By writing the time-reversal partner of $\epsilon(\bm{k})$
as $\epsilon_\Theta(-\bm{k})$, we obtain the BdG Hamiltonian in the band representation
\begin{subequations}\begin{align}
H_\B(\bm{k})&\equiv U_\B(\bm{k})H_{\mathrm{BdG}}(\bm{k})U_\B^\dagger(\bm{k})\\
&=\begin{pmatrix}\epsilon(\bm{k})&\Delta_\B(\bm{k})\\\Delta_\B(\bm{k})^\dagger&-\epsilon_\Theta(-\bm{k})
    \end{pmatrix},\label{eq:X_b_def}
\end{align}\end{subequations}
with $U_\B(\bm{k})=\diag(U_{\rm N}(\bm{k}),\Theta U_{\rm N}(-\bm{k})\Theta^{-1})$.
The order parameter in the band basis is given by
\begin{align}
    [\Delta_\B(\bm{k})]_{nm}&=\Delta_0\braket{u_n(\bm{k})|\Theta u_m(-\bm{k})},\label{eq:Delta_b}
\end{align}
with $H_{\rm N}(\bm{k})\ket{u_n(\bm{k})}=\epsilon_n(\bm{k})\ket{u_n(\bm{k})}$.

An important point here is that the system described by a generally complicated Bloch Hamiltonian $H_{\rm N}(\bm{k})$ recasts into another system with a simple normal-state Hamiltonian $\epsilon(\bm{k})$, a diagonal matrix.
The effective order parameter Eq.~\eqref{eq:Delta_b} generally has off-diagonal components but becomes diagonal in the presence of time-reversal symmetry.
Thus, in this case,
Eq.~\eqref{eq:X_b} recasts into the collection of the small 2-by-2 matrices of the form
\begin{align}
    h_{nn}(\bm{k})\equiv\begin{pmatrix}
    \epsilon_n(\bm{k})&\Delta_0\\
   \Delta_0^*&-\epsilon_n(\bm{k})
    \end{pmatrix}.\label{eq:2by2}
\end{align}
This allows us to intuitively understand the system based on the textbook knowledge of the single-band superconductivity: For example, the low-energy spectrum is given by $\pm\sqrt{\epsilon_n(\bm{k})^2+|\Delta_0|^2}$.

\subsection{Quantum geometry in Cooper pairs}
We have seen that the system may be understood by using the single-band model, in particular in the presence of time-reversal symmetry.
Indeed, the prescription is 
sufficient to capture spectral properties like the specific heat and density of states.
However, such identification of multiband systems with the assembly of single-band models is incomplete if we are to consider the {response} to the perturbation.

To be specific, let us consider the case of the Cooper-pair momentum $2\bm{q}$, or the applied supercurrent.
For a single-band $s$-wave superconductor, $\bm{q}$ is incorporated by
\begin{align}
    h_{\mathrm{single}}(\bm{k};\bm{q})=\begin{pmatrix}
    \epsilon(\bm{k}+\bm{q})&\Delta_0\\
    \Delta_0&-\epsilon(\bm{k}-\bm{q})
    \end{pmatrix}.
    \label{eq:single_q}
\end{align}
The $O(q)$ contribution causes the so-called Doppler shift of the energy dispersion: The spectrum shifts upward or downward by $\bm{q}\cdot\partial_{\bm{k}}\epsilon(\bm{k})$.
This effect makes the paramagnetic-current contribution to the superfluid weight, which is negligible at low temperatures.
The $O(q^2)$ correction gives rise to the diamagnetic-current contribution and is responsible for the conventional expression of the superfluid weight $D_{\mathrm S}^{ij}\simeq n_{\rm e}/m_{ij}$ with $n_{\rm e}$ the electron density and $1/m_{ij}$ the inverse effective mass tensor.

In contrast to the ideal single-band superconductivity, the $(n,n)$ sector of the band-represented Hamiltonian has the following form:
\begin{align}
    h_{nn}(\bm{k};\bm{q})=\begin{pmatrix}
    \epsilon_n(\bm{k}+\bm{q})&[\Delta_\B(\bm{k};\bm{q})]_{nn}\\
    [\Delta_\B(\bm{k};\bm{q})]_{nn}^*&-\epsilon_n(\bm{k}-\bm{q})
    \end{pmatrix}.\label{eq:hnn_k_q}
\end{align}
The effective order parameter in the band basis is given by
\begin{align}
    [\Delta_\B(\bm{k};\bm{q})]_{nn}&=\Delta_0\braket{u_n(\bm{k}+\bm{q})|u_n(\bm{k}-\bm{q})}\notag\\
    &=e^{-i\theta_{nn}(\bm{k},\bm{q})}\Delta_0(1-q_iq_j\QM_n^{ij}(\bm{k})),\label{eq:delta_g_intra}
\end{align}
up to $O(q^2)$.
Here, we defined the phase $\theta_{nm}(\bm{k},\bm{q})\equiv \gamma_n(\bm{k},\bm{q})-\gamma_m(\bm{k},-\bm{q})$, with $\gamma_n(\bm{k},\bm{q})=\int_0^{\bm{q}}d\bar{{q}_j}{A}^j_{nn}(\bm{k}+\bar{\bm{q}})=q_jA^j_{nn}(\bm{k})+O(q^2)$.
This is related to the Berry phase and is the Wilson line along the open straight path $0\to\bm{q}$.
Note that the Wilson line includes the Berry connection $A_{nm}^i(\bm{k})\equiv-i\braket{u_n(\bm{k})|\partial_{k_i}u_m(\bm{k})}$ and explicitly ensures the gauge covariance.
\footnote{
Indeed, the factor $e^{-i\theta_{nn}(\bm{k},\bm{q})}$ acquires $e^{-i[\chi_n(\bm{k}+\bm{q})-\chi_n(\bm{k}-\bm{q})]}$ upon gauge transformation $\ket{u_n(\bm{k})}\to\ket{u_n(\bm{k})}e^{i\chi_n(\bm{k})}$, which is the property $[\Delta_\B(\bm{k};\bm{q})]_{nn}$ should follow as understood from the first line of Eq.~\eqref{eq:delta_g_intra}.
Direct Taylor expansion of the first line of Eq.~\eqref{eq:delta_g_intra} will lead to collections of various gauge-dependent quantities, and thus the introduction of the Wilson line is quite helpful to obtain physically transparent expressions.}

The important point is the appearance of the quantum metric $\QM_n^{ij}(\bm{k})\equiv\sum_{m\neq n}A^i_{nm}(\bm{k})A^j_{mn}(\bm{k})+\text{c.c.}$ in the gap amplitude, Eq.~\eqref{eq:delta_g_intra}.
This quantity measures the ``distance," or the difference, of the neighboring Bloch states at $\bm{k}\pm\bm{q}$. Indeed, we obtain
\begin{align}
2q_iq_j\QM_n^{ij}(\bm{k})&= 1-|\braket{u_n(\bm{k}+\bm{q})|u_n(\bm{k}-\bm{q})}|^2\ge0,
\end{align}
up to $O(q^2)$, which vanishes for $\bm{q}=0$.
For an intuitive understanding, note that for $\bm{q}=0$, Cooper pairs are formed between the states with $\bm{k}$ and $-\bm{k}$, whose wave functions are essentially equivalent owing to the time-reversal symmetry.
In the presence of $\bm{q}\neq0$, there appears a mismatch of the wave functions with the wave numbers $\bm{k}+\bm{q}$ and $-\bm{k}+\bm{q}$, since they are not related by symmetry.
This mismatch causes difficulty in forming the pair, decreasing the amplitude of the effective pair potential.
Such a correction to the band-represented pair potential 
is dubbed QGPP in the following.

Another aspect of QGPP is the appearance of the interband component,
\begin{align}
    [\Delta_\B(\bm{k};\bm{q})]_{nm}=\Delta_0[-2iq_i A^i_{nm}(\bm{k})]e^{-i\theta_{nm}(\bm{k},\bm{q})},
    \label{eq:delta_g_inter}
\end{align}
for $n\neq m$ up to $O(q^2)$.
Equation~\eqref{eq:delta_g_inter} indicates that the interband component necessarily appears under finite $\bm{q}$, or the supercurrent, ensured by quantum geometry.
Thus, we cannot simply identify the system as the collection of 
{the single-band superconductor with $h_{\rm single}(\bm{k};\bm{q})$}, as the different bands $n$ and $m$ are coupled through Eq.~\eqref{eq:delta_g_inter}.

{In summary, combining Eqs.~\eqref{eq:delta_g_intra} and~\eqref{eq:delta_g_inter}, QGPP for the parameter $\bm{q}$ is defined by $\Delta_{\geom}(\bm{k};\bm{q})$ as follows:}
\begin{align}
[\Delta_\B(\bm{k};\bm{q})]_{nm}e^{i\theta_{nm}(\bm{k},\bm{q})}=\Delta_0\,\delta_{nm}+[\Delta_{\geom}(\bm{k};\bm{q})]_{nm},
\label{eq:delta_g_summary}
\end{align}
{while generalized definitions are given in latter sections by Eqs.~\eqref{eq:Delta_geom_general_pre} and~\eqref{eq:delta_geom_general_non_Abelian}.}
In particular, $\Delta_\geom(\bm{k};\bm{q})$ is $O(q^2)$ and $O(q)$ for the intra- and inter-band components, respectively.
The phase factor $e^{i\theta_{nm}(\bm{k},\bm{q})}$ can be removed by a unitary transformation and is not essential for thermodynamic properties.
Thus, the appearance of $\Delta_\geom(\bm{k};\bm{q})$ is the only and essential difference in the single- and multi-band models for thermodynamic properties.

{The expression of the intraband QGPP for the supercurrent in spin-singlet superconductors Eq.~\eqref{eq:delta_g_intra} has been pointed out in Ref.~\cite{Liang2017-nj} with allowing spatially-nonuniform pairings.
Here and hereafter in this paper, we give the unified understanding and description of QGPP for general superconducting order parameters with internal degrees of freedom such as sublattices and spin-triplet pairings.
}
Note {also} that expressions similar to Eqs.~\eqref{eq:delta_g_intra} and~\eqref{eq:delta_g_inter} in Refs.~\onlinecite{Villegas2021-gz,Villegas2023-gy} are not gauge covariant and thus different from QGPP.

\subsection{Geometric contribution to free energy and susceptibilities}
The presence of QGPP implies that multiband spin-singlet superconductors respond to the perturbation differently from the collection of single-band superconductors.
To see this, we expand the free energy $\Omega(\bm{q})$ in terms of the $O(q)$ quantity $\Delta_\geom(\bm{k};\bm{q})$, in addition to the $\bm{q}$ dependence of the normal-state dispersion.
We obtain up to $O(q^2)$,
\begin{align}
    \Omega(\bm{q})&=-\int\frac{d^dk}{(2\pi)^d}\frac{1}{2\beta}\tr\ln\left[1+e^{-\beta H_\B(\bm{k};\bm{q})}\right]+\text{const.}\notag\\
    &=\Omega(0)+\Omega_\conv(\bm{q})+\Omega_{\geom}(\bm{q}).\label{eq:temp_free_energy}
\end{align}
Note that we can use $H_{\B}(\bm{k};\bm{q})$ instead of $H_{\BdG}(\bm{k};\bm{q})$ to evaluate the free energy according to the circularity of the trace.
The conventional free energy $\Omega(0)+\Omega_c(\bm{q})$ is calculated by the first line of Eq.~\eqref{eq:temp_free_energy}
with replacing $\Delta_\B(\bm{k};\bm{q})$ with $\Delta_0$.
In this contribution, the effect of $\bm{q}$ is incorporated only through the normal-state energy dispersion, and therefore, single-band results hold true.
Accordingly, $\Omega_c(\bm{q})$ is 
divided into paramagnetic- and diamagnetic-current contributions,
\begin{align}
    \Omega_\conv(\bm{q})=\Omega_{\para}(\bm{q})+\Omega_{\dia}(\bm{q}),
\end{align}
only the latter of which survives at zero temperature.
On the other hand, QGPP contribution to the free energy $\Omega_\geom(\bm{q})$ describes the correction by $\Delta_\geom(\bm{k};\bm{q})$ to the single-band picture.
After some calculations in Appendix~\ref{app:quantum-geometric_expansion}, $\Omega_\para(\bm{q})$, $\Omega_\dia(\bm{q})$ and 
$\Omega_\geom(\bm{q})$ are written down as follows:
\begin{widetext}
\begin{gather}
\Omega_\para(\bm{q})=\frac{1}{2}q_iq_j\sum_{n}\int_\BZ\frac{d^dk}{(2\pi)^d}\partial_{q_i}\epsilon_n\partial_{q_j}\epsilon_n\,f'(E_n),
\quad\Omega_\dia(\bm{q})=\frac{1}{4}q_iq_j\sum_n\int_\BZ\frac{d^dk}{(2\pi)^d}\partial_{q_i}\partial_{q_j}\epsilon_n\left[1-\frac{\epsilon_n}{E_n}\tanh\frac{\beta E_n}{2}\right],\notag\\
    \Omega_\geom(\bm{q})=\frac{1}{8}q_iq_j\int_\BZ\frac{d^dk}{(2\pi)^d}\Delta_0^2\sum_{n\neq m}\QM_{nm}^{ij}\sum_{s,s'=\pm1}\frac{(\epsilon_n-\epsilon_m)^2}{E_nE_m}ss'\frac{f(sE_n)-f(s'E_m)}{sE_n-s'E_m}.
    \label{eq:Omega_all}
\end{gather}
\end{widetext}
Here, $\partial_{\bm{q}}$ indicates $\lim_{q\to0}\partial_{\bm{q}}\epsilon_n(\bm{k}+\bm{q})=\partial_{\bm{k}}\epsilon_n(\bm{k})$, for example.
We defined the quasiparticle energy $E_n=\sqrt{\epsilon_n^2+\Delta_0^2}$ and the band-resolved quantum metric $\QM_{nm}^{ij}=A^{i}_{nm}A^{j}_{mn}+\text{c.c.}$, while the wave-number dependence of quantities is implicit.
The integral runs over the first Brillouin zone (BZ), and $d$ is the dimension of the system.
The Fermi distribution function is denoted by $f(E)=(e^{\beta E}+1)^{-1}$ with the inverse temperature $\beta$.

The $O(q^2)$ coefficient of the free energy $\Omega(\bm{q})$ describes the Meissner effect, and is known as the superfluid weight $D_{\mathrm S}^{ij}$.
According to Eq.~\eqref{eq:Omega_all}, $D_{\mathrm S}^{ij}$ is given by the sum of three pieces: In addition to the text-book paramagnetic- and diamagnetic-current contributions $D^{ij}_\para+D^{ij}_\dia$, there exists QGPP contribution to the superfluid weight unique to multiband systems $D^{ij}_\geom$:
\begin{align}
    D_{\mathrm S}^{ij}=D^{ij}_\para+D^{ij}_\dia+D^{ij}_\geom,
\end{align}
each of which is naturally given by $D^{ij}_\geom=\partial_{q_i}\partial_{q_j}\Omega_\geom(\bm{q})$ and so on.
The contribution from QGPP, $D_{\geom}^{ij}$,
reproduces the geometric superfluid weight in the literature~\cite{Peotta2015-tv,Liang2017-ax} {(see also Appendix~\ref{app:quantum-geometric_expansion} and Eq.~\eqref{eq_supp:Omega_1})}, and thus QGPP {offers a way to interpret it.}

\subsection{
{Generalization to other perturbations}
}
The concept of QGPP
can be generalized to any perturbation of the normal state.
Equations~\eqref{eq:hnn_k_q}-\eqref{eq:Omega_all} remain valid for an arbitrary time-reversal-breaking parameter by simply replacing $\bm{q}$.
For the case of the Zeeman magnetic field $\bm{h}$, we replace $q_i$ with $h_i$,
and accordingly, the Berry connection is replaced with that for the field $h_i$,
\begin{align}
    A^{h_i}_{nm}(\bm{k})&\equiv-i\braket{u_n(\bm{k};\bm{h})|\partial_{h_i}u_m(\bm{k};\bm{h})}|_{\bm{h}\to0}
    \notag\\&=\frac{\braket{u_n(\bm{k})|s^i|u_m(\bm{k})}}{\epsilon_n(\bm{k})-\epsilon_m(\bm{k})}\quad (n\neq m),
\end{align}
which describes the quantum geometry related to spin $\bm{s}$.
On the other hand, QGPP does not appear under time-reversal-symmetric perturbations for the case of spin-singlet superconductivity in systems of non-degenerate bands.
Indeed, the wave function of an electron remains essentially the same as its time-reversal partner, and thus $\Delta_\B=\Delta_0 {\bm{1}}$ is preserved according to Eq.~\eqref{eq:Delta_b}.
The absence of QGPP reminds us of the absence of the depairing by time-reversal-symmetric perturbations known as the Anderson's theorem~\cite{Anderson1959-xf}, although
QGPP is absent not only for the $s$-wave but also for the $d$-wave superconductivity as is clear by replacing $\Delta_0\to \Delta_0(k_x^2-k_y^2)$.

The generalization allows us to describe the crossed response of $\bm{q}$ and $\bm{h}$ as well.
Let us write time-reversal-breaking fields as $\bm{X}=(\bm{q},\bm{h})$.
The generalized susceptibility reads
\begin{align}
\chi^{X_aX_b}\!\equiv\partial_{X_a}\partial_{X_b}\Omega(\bm{X})=\chi^{X_aX_b}_\para+\chi^{X_aX_b}_\dia\!+\chi^{X_aX_b}_\geom\!.
\end{align}
This corresponds to the superfluid weight for $(X_a,X_b)=(q_i,q_j)$ and spin susceptibility for $(X_a,X_b)=(h_i,h_j)$, while describes the supercurrent-induced magnetization, i.e. the superconducting Edelstein effect~\cite{Edelstein1995-mf,He2019-bi,He2020-yq,Ikeda2020-om,He2021-ra,Chirolli2022-mz}, for $(X_a,X_b)=(h_i,q_j)$.
$\Omega(\bm{X})$ and thus $\chi^{X_aX_b}$ are immediately obtained by replacing $\bm{q}\to\bm{X}$ in Eqs.~\eqref{eq:Omega_all}, in particular with $\QM_{nm}^{ij}\to \QM_{nm}^{X_aX_b}\equiv A_{nm}^{X_a}A_{mn}^{X_b}+\text{c.c.}$.
The QGPP correction of the susceptibility also implies the change in the transition temperature, as expected from $\Omega_\geom(\bm{X})\propto\Delta_0^2$.

To illustrate the QGPP contribution to the susceptibilities, let us consider the two-band model 
\begin{align}
H_{\rm N}(\bm{k}+\bm{q};\bm{h})=\xi(\bm{k}+\bm{q})+[\bm{g}(\bm{k}+\bm{q})-\bm{h}]\cdot\bm{s},\label{eq:twobandmodel}
\end{align}
{where the first term is proportional to the identity matrix.}
This is a minimal model for the noncentrosymmetric superconductors under the supercurrent and the magnetic field~\cite{Bauer2012-xi,Smidman2017-hb}.
Here, $\bm{g}(\bm{k})=-\bm{g}(-\bm{k})$ represents the antisymmetric spin-orbit coupling, which is allowed without the inversion symmetry.

At zero temperature, we obtain $\chi_{\para}^{X_iX_j}=0$ and
\begin{subequations}\begin{align}
\chi^{X_iX_j}_{\dia}&=\int_\BZ\frac{d^dk}{(2\pi)^d}\sum_{n=\pm1}\frac{1}{m^{X_iX_j}_n}\,\frac{1}{2}\left[{1}-\frac{\epsilon_n}{E_n}\right],\label{eq:chi_dia_twoband}\\
\chi_\geom^{X_iX_j}&=4\int_\BZ\frac{d^dk}{(2\pi)^d}\,\QM^{X_iX_j}\frac{g^2\Delta_0^2}{(E_++E_-)E_+E_-},\label{eq:chi_g_twoband}
\end{align}\end{subequations}
from Eqs.~\eqref{eq:Omega_all}
with $\epsilon_n=\xi+ng$, $g=|\bm{g}|$, and the helicity $n=\pm1$.
Here, the $\bm{k}$ dependence is implicit.
The susceptibilities are determined by the generalized inverse mass tensor $1/m_n^{X_iX_j}\equiv\partial_{X_i}\partial_{X_j}\epsilon_n$ and quantum metric tensor $\QM^{X_iX_j}$, which is given by
\begin{align}
   \! \!\! \begin{pmatrix}\QM^{q_iq_j}&\QM^{q_iH_j}\\ \QM^{H_iq_j}&\QM^{H_iH_j}
    \end{pmatrix}\!&=\begin{pmatrix}
    \frac{1}{2}\partial_{k_i}\hat{g}\cdot\partial_{k_j}\hat{g}&-\frac{1}{2g}\partial_{k_i}\hat{g}_j\\-\frac{1}{2g}\partial_{k_j}\hat{g}_i&\frac{1}{2g^2}(\delta_{ij}-\hat{g}_i\hat{g}_j)
    \end{pmatrix}.
\end{align}
The direction of the g-vector is denoted by $\hat{g}\equiv\bm{g}/g$.

The susceptibility is further simplified for weak-coupling superconductors, where the Fermi energy $E_{\rm F}$ is much larger than the g-vector $g$ and the order parameter $\Delta_0$.
We obtain
\begin{subequations}
\begin{align}
    \chi^{X_iX_j}_\dia&\simeq[\chi^{X_iX_j}_\dia]_{\Delta_0=0},\label{eq:chi_dia_deepband}\\
    \chi_\geom^{X_iX_j}&\simeq 4N(0)\Braket{g^2\,\QM^{X_iX_j}I(\Delta_0/g)}_{\FS},
\end{align}\label{eq:chi_deep_band}\end{subequations}
\begin{figure}
    \centering
    \includegraphics[width=0.3\textwidth]{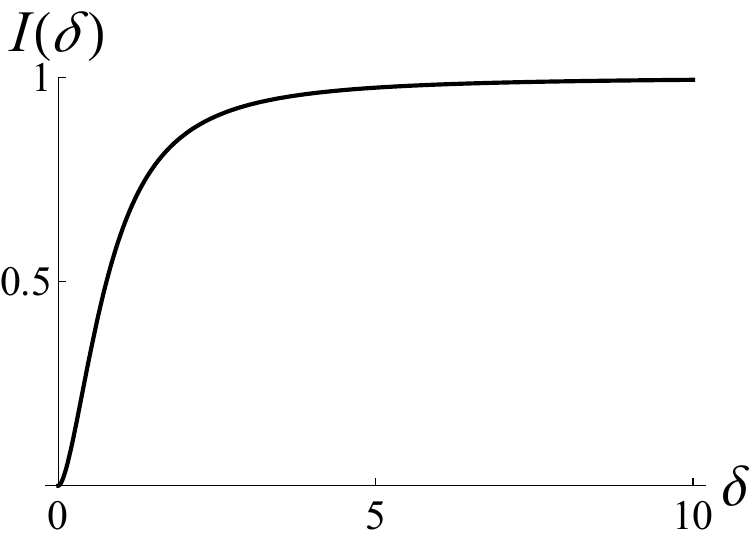}
    \caption{Plot of the function $I(\delta)\equiv\frac{\delta^2}{\sqrt{1+\delta^2}}\tanh^{-1}\frac{1}{\sqrt{1+\delta^2}}$.}
    \label{fig:I_of_x}
\end{figure}
with the Fermi-level density of states $N(0)$ defined for $\xi(\bm{k})$ and the average over its Fermi surface $\braket{\cdot}_{\FS}$.
As is the case for the superfluid weight, the diamagnetic contribution
$\chi_{\dia}^{X_iX_j}$ is approximated by that in the normal state.
The geometric contribution includes 
 a dimensionless function $I(\delta)\equiv\frac{\delta^2}{\sqrt{1+\delta^2}}\tanh^{-1}\frac{1}{\sqrt{1+\delta^2}}$
, which {is obtained after integration over $\xi$ and} monotonically increases from $0$ to $1$ as $\delta=\Delta_0/g$ grows [Fig.~\ref{fig:I_of_x}].
We neglected the $\xi$ dependence of the $g$-vector, the generalized inverse mass tensor, and the quantum metric, for simplicity.

For the case of the superfluid weight $(X_a,X_b)=(q_i,q_j)$, Eqs.~\eqref{eq:chi_dia_twoband} and~\eqref{eq:chi_dia_deepband} clearly reproduces the Fermi-liquid formula $D^{ij}_\conv\sim n_{\rm e}/m^{ij}$.
The ratio of the geometric term to $D_{\rm c}^{ij}$
is estimated to be
${\chi^{q^iq^j}_\geom}/{\chi_\dia^{q_iq^j}}\sim \frac{g^2}{E_{\rm F}^2}I(\delta)$, that is, $\sim\frac{\Delta^2_0}{E_{\rm F}^2}\ln\left[\frac{g}{\Delta_0}\right]$
for ${\Delta_0} \ll g$, and $\sim g^2/E_{\rm F}^2$ for $g\ll {\Delta_0}$.
Thus, the geometric superfluid weight would become important in the BCS-BEC crossover regime $E_{\rm F}\sim\Delta$ as exemplified in previous studies~\cite{Peotta2015-tv,Liang2017-ax,Rossi2021-hk,Huhtinen2022-mp,Torma2022-gd}, though strictly speaking the weak-coupling expressions~\eqref{eq:chi_deep_band} are not applicable in the strong-coupling region $E_{\rm F}\sim\Delta$.

For the spin susceptibilities $(X_a,X_b)=(h_i,h_j)$, 
only the information around the Fermi surfaces comes into play due to the cancellation between different helicity bands.
We obtain
\begin{subequations}
    \begin{align}
    \chi_{\dia}^{h_ih_j}&\simeq -2N(0)\braket{\delta_{ij}-\hat{g}_i\hat{g}_j}_{\FS},\\
    \chi_{\geom}^{h_ih_j}&\simeq 2N(0)\braket{I(\Delta_0/g)(\delta_{ij}-\hat{g}_i\hat{g}_j)}_{\FS}.
\end{align}\end{subequations}
Here, an extra minus sign appears compared with the usual definition $\partial M_i/\partial h_j=-\chi^{h_ih_j}$.
Note that the spin susceptibility $\chi_\dia^{h_ih_j}+\chi_\geom^{h_ih_j}$ vanishes in the centrosymmetric spin-singlet superconductors ($g\ll\Delta_0)$, while only the $\hat{h}\parallel\hat{g}$ part vanishes in the noncentrosymmetric limit $(g\gg\Delta_0)$, as it should be~\cite{Smidman2017-hb,Bauer2012-xi}.
It should be noticed that QGPP is responsible for the crossover between these two limits, whose key parameter is $\Delta_0/g$ instead of $\Delta_0/E_{\rm F}$.
{Note also that for $g\neq0$, $\chi_{\rm dia}^{h_ih_j}$ corresponds to the Van-Vleck susceptibility of the normal state, while the Pauli susceptibility vanishes in the superconducting state at zero temperature. The geometric susceptibility $\chi_{\geom}^{h_ih_j}$ makes a superconducting and quantum-geometric correction to the Van-Vleck term. This is consistent with the observation in Ref.~\cite{Iskin2018-as}.}

The superconducting Edelstein effect is described by the response formula $M_i=\alpha_{ij}j_j$, with magnetization $\bm{M}$ and supercurrent $\bm{j}$.
The coefficient $\alpha_{ij}=\chi^{h_iq_k}[\chi^{qq}]^{-1}_{kj}$ depends on the susceptibility $\chi^{h_iq_j}$, which is obtained as
\begin{subequations}
\begin{align}
    \chi_{\dia}^{h_iq_j}&\simeq 2N(0)\braket{g\partial_{k_j}\hat{g}_i}_{\FS},\\
\chi_\geom^{h_iq_j}&\simeq-2N(0)\braket{gI(\Delta_0/g)\partial_{k_j}\hat{g}_i}_{\FS}.
\end{align}\end{subequations}
The ratio of QGPP contribution to conventional one is again $I(\Delta_0/g)\sim O(\Delta_0/g)^2$ for $\Delta_0\ll g$, and thus $\chi_\geom^{h_iq_j}$ is quantitatively important when $\Delta_0\sim g$.

In summary, the QGPP contribution 
can be important when $\Delta_0$ is comparable to some of the energy scales in the normal state.
For the superfluid weight, a large ratio of the order parameter to Fermi energy is required, and thus the BCS-BEC superconductors such as FeSe~\cite{Shibauchi2020-gj}, Li$_x$ZrNCl~\cite{Kasahara2015-sn,Nakagawa2018-cw,Nakagawa2021-ka}, and twisted bi- and tri-layer graphene~\cite{Cao2018-hg,Park2021-kk,Kim2022-kw,Chen2024-do} offer a good platform.
For the spin susceptibility and the superconducting Edelstein coefficient, the ratio to spin-orbit coupling instead of Fermi energy is important.
In particular, QGPP plays an essential role in the crossover from the centrosymmetric to noncentrosymmetric spin-singlet superconductivity.

In this section, we have focused on the quantum-geometric terms in the free energy quadratic regarding the external fields.
On the other hand, quantum-geometric terms can also appear in the first order of the external fields.
Anapole superconductivity~\cite{Kanasugi2022-cl} offers such an example, which refers to the superconductivity spontaneously breaking the time-reversal symmetry but preserving the $\Theta I$ symmetry{, namely the combined time-reversal and inversion symmetry}.
More precisely, the {quadratic product} of the order parameter should have a component symmetry-equivalent to the supercurrent: This defining property allows the $\bm{q}$-linear coupling in the free energy.
The quantum-geometric origin of such a coupling has been discussed~\cite{Kitamura2023-on}, and would also be understood based on the concept of QGPP.
The impact of quantum geometry on field-linear couplings in exotic superconductivity is an interesting future direction.

\subsection{Generalization to non-s-wave states}\label{sec:non-s-wave}
We here generalize QGPP to arbitrary pair potentials for non-degenerate bands and discuss its
potential ability to engineer exotic superconducting phenomena.
Generally speaking, pair potential $\Delta(\bm{k})$ has a matrix structure and depends on the wave number.
For example, we can write $\Delta(\bm{k})=\psi(\bm{k})+\bm{d}(\bm{k})\cdot\bm{s}$ when the system has only the spin degree of freedom.
Here, $\psi(\bm{k})$ and $\bm{d}(\bm{k})$ describe the spin-singlet and -triplet Cooper pairs, respectively, and coexist in the absence of the inversion symmetry.
The pair potentials in the band basis with and without time-reversal breaking fields $\bm{X}$ are then given by $[\Delta_\B(\bm{k})]_{nm}=\braket{u_n(\bm{k})|\Delta(\bm{k})|u_m(\bm{k})}$ and 
\begin{align}
[\Delta_\B(\bm{k};\bm{X})]_{nm}&=\braket{u_n(\bm{k};\bm{X})|\Delta(\bm{k})|u_m(\bm{k};-\bm{X})},\label{eq:Delta_b_general}
\end{align}
respectively, instead of Eqs.~\eqref{eq:Delta_b},~\eqref{eq:delta_g_intra} and~\eqref{eq:delta_g_summary}{, for arbitrary systems without degenerate bands}.

QGPP induced by time-reversal breaking fields $\bm{X}$ can be obtained by expanding $\Delta_\B(\bm{k};\bm{X})$.
In doing so, it is transparent to expand the combination
$[\Delta_\B(\bm{k};\bm{X})]_{nm}e^{i\theta_{nm}(\bm{k};\bm{X})}$
rather than $[\Delta_\B(\bm{k};\bm{X})]_{nm}$ itself to ensure the gauge covariance, as detailed in Appendix~\ref{app:Delta_g_derivation}.
The result is 
\begin{align}
\!\!\![\Delta_\B(\bm{k};\bm{X})]_{nm}e^{i\theta_{nm}(\bm{k};\bm{X})}=[\Delta_\B(\bm{k})+\Delta_\geom(\bm{k};\bm{X})]_{nm},\label{eq:Delta_geom_general_pre}
\end{align}
with
\begin{align}
    \Delta_\geom(\bm{k};\bm{X})=-iX_i\{A_\inter^{X_i}(\bm{k}),\Delta_\B(\bm{k})\}+O(X^2).
    \label{eq:Delta_geom_general}
\end{align}
Here, we defined the interband Berry connection $[A^{X_i}_\inter(\bm{k})]_{nm}\equiv(1-\delta_{n,m})A_{nm}^{X_i}(\bm{k})$. 
Similarly to the $s$-wave case, Eq.~\eqref{eq:Delta_geom_general} ensures the presence of the interband order parameter under time-reversal breaking fields such as supercurrent and Zeeman field.

Interestingly, a finite intraband component can be obtained from Eq.~\eqref{eq:Delta_geom_general} when the original order parameter has an interband component.
Let us consider the system with only the spin degree of freedom and lacks inversion symmetry.
The normal-state Hamiltonian is generally written as {a two-by-two matrix,}
\begin{align}
H_{\rm N}(\bm{k})=\xi(\bm{k})+\bm{g}(\bm{k})\cdot\bm{s},
\end{align}
where $\bm{g}(\bm{k})$ is the anti-symmetric spin-orbit coupling such as the Rashba-type one.
For the pair potential $\Delta(\bm{k})=\psi(\bm{k})+\bm{d}(\bm{k})\cdot\bm{s}$,
the intraband component without the field $\bm{X}$ is given by $[\Delta_\B(\bm{k})]_{nn}=\psi(\bm{k})+n\bm{d}(\bm{k})\cdot\hat{g}(\bm{k})$, where $n=\pm1$ is the band index or the helicity of normal electrons and specifies electron's spin parallel and antiparall to the g-vector, respectively.
The intraband component of QGPP induced by the Zeeman field $-\bm{h}\cdot\bm{s}$ is given by
\begin{align}
[\Delta_\geom(\bm{k};\bm{h})]_{nn}&=\frac{i}{|\bm{g}(\bm{k})|^2}\bm{h}\cdot\bm{g}(\bm{k})\times\bm{d}(\bm{k}).\label{eq:parainducedgap}
\end{align}
For the case of the supercurrent $\bm{X}=\bm{q}$, the Zeeman field $\bm{h}$ on the right-hand side is replaced with $-[\bm{q}\cdot\nabla_{\bm{k}}]\bm{g}(\bm{k})$.
The derivation of the results are given in Appendix~\ref{app:QGPP_for_the_two-band_model}.

Importantly, the induced QGPP inherits the symmetry of the external field $\bm{X}$: 
The effective intraband pair potential 
$[\Delta_\B(\bm{k})+\Delta_\geom(\bm{k};\bm{X})]_{nn}$ 
breaks time-reversal symmetry and is analogous to that of chiral superconductivity.
This implies the possibility of TSC induced by quantum geometry.
Indeed, Eq.~\eqref{eq:parainducedgap} gives a unified description of the previously proposed gap opening and TSC in mixed $d$- and $p$-wave superconductors under the Zeeman field~\cite{Daido2016-xk,Daido2017-ij,Yanase2022-pw} and supercurrent~\cite{Takasan2022-pq,Sumita2022-ul}.
With Rashba-type spin-orbit coupling and Zeeman field, the intraband pair potential is of the form [see Ref.~\cite{Yanase2022-pw} for deitals]
\begin{align}
[\Delta_\B(\bm{k})+\Delta_\geom(\bm{k};\bm{X})]_{nn}\sim k_x^2-k_y^2+i\,k_xk_y.
\end{align}
This effectively realizes spinful chiral $d_{x^2-y^2}+id_{xy}$-wave superconductivity with Chern number $4$.

It should be noted that QGPP is not the only way to encode the quantum geometry of normal states into Bogoliubov quasiparticles.
Actually, the unitary matrix to band representation $U_\B(\bm{k})$ can contribute to topological invariants, though not to the free energy and thus not to thermodynamic properties.
A finite Chern number of topological $s$-wave superconductivity~\cite{Sato2009-cn,Sato2010-qj} is achieved by its contribution.
In this sense, the QGPP contribution to TSC captures the topological properties relative to the putative system where the effective order parameter in the band basis $\Delta_\B(\bm{k};\bm{X})=\Delta_\B(\bm{k})+\Delta_\geom(\bm{k};\bm{X})$ is formally replaced with a trivial $s$-wave order parameter $\Delta_0\bm{1}$.
The engineering of exotic superconducting properties based on both $U_\B(\bm{k})$ and QGPP is important especially to identify TSC candidates.
The former is completely determined by the normal-state properties, while the effect of nontrivial pair potential is fully captured by QGPP.

\section{QGPP in systems with entangled bands}
\label{sec:secIII}
We finally discuss the general situations where the bands can have nontrivial band degeneracy.
This allows us to describe systems with combined inversion and time-reversal symmetries along with the spin-orbit coupling, {such as centrosymmetric metals at zero magnetic fields and $\Theta I$-symmetric antiferromagnets.}
The formalism can also be applied to several bands that are entangled and have degeneracy in some discrete points in the Brillouin zone, {that is, Dirac and Weyl fermions in the general sense.}

\subsection{QGPP and generalized band representation}
Here we show the results of QGPP and related concepts for general situations.
The derivation is given by introducing the non-Abelian generalization of the phase factor $e^{i\theta_{nm}(\bm{k};\bm{X})}$ and also the concept of the quantum-geometry factor, as discussed in Appendix~\ref{app:derivation_general}.

We can generally perform a unitary transformation of the BdG Hamiltonian to obtain the expression like
\begin{align}
\bar{H}_\B(\bm{X})
&=
\begin{pmatrix}\bar{\epsilon}(\bm{X})&\Delta_\B+\Delta_\geom(\bm{X})\\
\Delta_\B^\dagger+\Delta_\geom(\bm{X})^\dagger&-\bar{\epsilon}(-\bm{X})
\end{pmatrix}.
\label{eq:GBR}
\end{align}
Here and hereafter, we drop the argument $\bm{k}$ unless necessary since all the quantities share the same wave number.
We call {Eq.~\eqref{eq:GBR}} the generalized band representation (GBR) of the BdG Hamiltonian.
{GBR recasts to the Bloch basis adopted to study nonlinear optical responses in Ref.~\cite{Sipe1993-rj} when $\bm{X}$ is the vector potential and the band degeneracy is absent.}
Let us consider the degenerate bands in the absence of the external fields.
The normal-state component $\bar{\epsilon}(\bm{X})$ is a block-diagonal matrix $[\bar{\epsilon}(\bm{X})]_{n\lambda,n'\lambda'}=\delta_{nn'}[\epsilon_n(\bm{X})]_{\lambda\lambda'}$, and is explicitly given by
\begin{align}
[\bar{\epsilon}_n(\bm{X})]_{\lambda\lambda'}&=\epsilon_n\delta_{\lambda\lambda'}+X_i\braket{u_n^{(\lambda)}|\partial_{X_i}H_{\rm N}|u_n^{(\lambda')}}\label{eq:energy}, 
\end{align}
neglecting $O(X^2)$ terms, whose leading-order term is available in Appendix~\ref{app:derivation_general}.
Here, we introduced the eigenstates $\ket{u_n^{(\lambda)}}$ for $\bm{X}=0$, whose eigenvalues are degenerate:
\begin{align}
H_{\rm N}\ket{u_n^{(\lambda)}}=\epsilon_n\ket{u_n^{(\lambda)}}\quad (\lambda=1,2,\cdots).\label{eq:Eq26}
\end{align}

We can show that the effective order parameter in GBR
\begin{align}
\bar{\Delta}_\B(\bm{X})\equiv \Delta_\B+\Delta_\geom(\bm{X}),\label{eq:delta_geom_general_non_Abelian}
\end{align}
is given by the sum of that in the absence of the field $\bm{X}$,
\begin{align}
[\Delta_\B]_{n\lambda,n'\lambda'}=\braket{u_n^{(\lambda)}|\Delta|u_{n'}^{(\lambda')}},
\end{align}
and QGPP,
\begin{align}
\Delta_\geom(\bm{X})&=-iX_i\{A^{X_i}_{{\rm inter}},\Delta_\B\}+\frac{1}{2}X_iX_j\Bigl(-i[A^{X_i,}_{{\rm inter};X_j},\Delta_\B]\notag\\
&\quad\quad -\{A^{X_i}_{{\rm inter}},\{A^{X_j}_{{\rm inter}},\Delta_\B\}\}\Bigr)+O(X^3),\label{eq:Delta_g_general}
\end{align}
whose first and third terms naturally generalize Eq.~\eqref{eq:Delta_geom_general} and Eq.~\eqref{eq:delta_g_intra}.
The second term includes the covariant derivative of the Berry connection $A^{X_i}_{\inter;X_j}$, which is a quantum-geometric quantity giving rise to the shift-current optical responses~\cite{Von_Baltz1981-kd,Sipe2000-wk,Cook2017-hh,Morimoto2023-ou}. 
It should also be noted that the last term does not coincide with the quantum metric except for the case of plain spin-singlet superconductivity $\Delta_\B\propto \bm{1}$.
The formula~\eqref{eq:Delta_g_general} is one of the central results of this paper, suggesting quantum-geometric corrections to equilibrium properties beyond the quantum metric.

The above expressions almost remain valid even when the degeneracy is slightly lifted, that is, for a system of entangled bands.
In this case, we have {only} to replace $\epsilon_n\to\epsilon_{n\lambda}$ in Eqs.~\eqref{eq:energy} and~\eqref{eq:Eq26}. 
We also note that similar expressions are obtained for time-reversal-symmetric perturbations.
We discuss this point in Appendix~\ref{app:QGPP_by_TRSP} and
show in Appendix~\ref{app:weaklynoncentro} an application to the weakly-noncentrosymmetric bilayer under the Zeeman field, where a noncentrosymmetric two-band model is extracted as the effective Hamiltonian.

{Note that we have neglected the external-field dependence of the pair potential in the orbital basis $\Delta$ to simplify the discussion. 
If it is necessary, we can simply replace the orbital-basis order parameter $\Delta$ with that determined self-consistently, namely $\Delta(\bm{X})$, by solving the gap equation in the external field.
This amounts to replacing $\Delta_\B$ here and hereafter with $\Delta_\B(\bm{X})$.
The net thermodynamic coefficients, for example, can be obtained by additionally expanding $\Delta_\B(\bm{X})$ by $\bm{X}$ up to $O(X^2)$.
If this is done, the obtained superfluid weight, for example, should coincide with that of e.g., Ref.~\cite{Huhtinen2022-mp}.
}

{As illustrated in previous sections for the plain $s$-wave superconductivity, QGPP describes all the quantum-geometric effects on the thermodynamic responses for a given pair potential in the orbital basis. 
This can be explicitly seen by following the discussion near Eq.~\eqref{eq:temp_free_energy} and that in Appendix~\ref{app:quantum-geometric_expansion} for the plain $s$-wave superconductivity.
The free energy in the field $\bm{X}$ is given by
\begin{align}
\Omega(\bm{X})&=-\int\frac{d^dk}{(2\pi)^d}\frac{1}{2\beta}\sum_{\omega_n}\tr\ln[\bar{G}_\B^{-1}(i\omega_n,\bm{X})]+\text{const.,}
\end{align}
where $\omega_n$ is the fermion Matsubara frequency and $\bar{G}_\B^{-1}(i\omega_n,\bm{X})=i\omega_n-\bar{H}_\B(\bm{X})$ is the inverse Green's function in GBR.
The inverse Green's function can be separated into two parts, i.e., the conventional one
\begin{align}
\bar{G}_\conv^{-1}(i\omega_n,\bm{X})=i\omega_n-\begin{pmatrix}
    \bar{\epsilon}(\bm{X})&\Delta_\B\\
    \Delta_\B^\dagger&-\bar{\epsilon}(-\bm{X})
\end{pmatrix}
\end{align}
and the QGPP in the Nambu space,
\begin{align}
\hat{\Delta}_\geom(\bm{X})\equiv\begin{pmatrix}
    0&\Delta_\geom(\bm{X})\\\Delta_\geom^\dagger(\bm{X})
\end{pmatrix}.
\end{align}
Since $\bar{G}_\conv^{-1}(i\omega_n,\bm{X})$ contains the $\bm{X}$ dependence only through the ``energy dispersion" $\bar{\epsilon}(\bm{X})$, 
$\bar{G}_\conv^{-1}(i\omega_n,\bm{X})$ contains no quantum-geometric effects in this sense.
\footnote{
{
In general $\bar{\epsilon}(\bm{X})$ can have a matrix structure, e.g., when the Kramers degeneracy is lifted by the magnetic field $\bm{h}$ as seen in the following sections.
In this case $\bar{\epsilon}(\bm{X})$ is the collection of two-by-two matrices. 
Each two-by-two sector describes the well-known anisotropic spin susceptibility of spin-triplet superconductors, for example.
In this paper, the quantum-geometric effects are intended to mean the effects beyond such situations.
}}
By using
\begin{align}
&\ln[\bar{G}_\B^{-1}(i\omega_n,\bm{X})]=\ln[\bar{G}_\conv^{-1}(i\omega_n,\bm{X})]\\
&\qquad+\ln[1-\bar{G}_\conv(i\omega_n,\bm{X})\hat{\Delta}_\geom(\bm{X})],
\end{align}
the first term describes the thermodynamic response without the quantum geometry, namely the conventional contribution.
It is now evident that all the quantum-geometric effects on the thermodynamic responses come from the second term, namely, the contribution from the QGPP.
}

In the remainder of this section, we illustrate the concept of QGPP and GBR in non-Abelian situations, taking a locally-noncentrosymmetric bilayer model with Kramers degeneracy as an example. 
Systems under the supercurrent and the Zeeman field are discussed order by order.

\subsection{Illustration with spinful bilayer models}
Let us consider a locally-noncentrosymmetric bilayer  
\begin{align}
H_{\rm N}(\bm{k})=\xi(\bm{k})+t_\perp(\bm{k})\eta_x+\bm{g}(\bm{k})\cdot\bm{s}\,\eta_z
\label{eq:locallynoncentro}
\end{align}
to illustrate QGPP.
Here, $\eta_i$ are the Pauli matrices in the layer space.
The Hamiltonian preserves the inversion and time-reversal symmetries by assuming $\bm{g}(\bm{k})=-\bm{g}(-\bm{k})$,
\begin{align}
IH_{\rm N}(\bm{k})I^\dagger=H_{\rm N}(-\bm{k}),\ \Theta H_{\rm N}(\bm{k})\Theta^{-1}=H_{\rm N}(-\bm{k}),
\end{align}
with $I=\eta_x$, $\Theta=is_yK$, and the complex-conjugation operator $K$.
Thus, the system has the Kramers degeneracy on the whole Brillouin zone.
The spectral decomposition of the Hamiltonian is given by
\begin{align}
H_{\rm N}=\sum_{n=\pm}\epsilon_nP_n,
\end{align}
with $\epsilon_n=\xi+nR$, $R= \sqrt{t_\perp^2+g^2}$, $g=|\bm{g}|$ and
\begin{align}
P_n&=\frac{n(H_{\rm N}-\epsilon_{-n})}{\epsilon_+-\epsilon_-}.
\end{align}
Here and hereafter, we abbreviate the argument $\bm{k}$ again.

To gain physical insight, it is convenient to adopt the manifestly-covariant Bloch basis (MCBB) for gauge fixing, because
MCBB makes the Kramers degree of freedom to transform in the same way as the real spin~\cite{Fu2015-xu}. 
For the case of the spinful bilayer models, the matrix element of an arbitrary matrix $A$ in the MCBB gauge can be calculated by the formula [see Appendix~\ref{app:MCBB}]
\begin{align}
&\braket{u_n^{\sigma}|A|u_{n'}^{\sigma'}}=[a^0_{nn'}(A)+\bm{a}_{nn'}(A)\cdot\bm{\sigma}]_{\sigma\sigma'},\notag\\
&a^\mu_{nn'}(A)=\frac{\Tr[P_+^\eta s^\mu P_nAP_{n'}]}{\sqrt{\Tr[P_nP_+^\eta]\Tr[P_{n'}P_+^\eta]}},\ \, P_+^\eta\equiv\frac{1+\eta_x}{2},\label{eq:formulas_me_A}
\end{align}
with $\mu=0,1,2,3$ and $s^0=1$.
Here we used $\sigma$ instead of $\lambda$ to distinguish the degenerate eigenstates, to emphasize that the degeneracy is due to the Kramers degree of freedom, i.e., the {pseudospin}.

\subsubsection{Zeeman field}
Let us consider the case of the Zeeman field.
In this case, $X_i\partial_{X_i}H_{\rm N}=-\bm{h}\cdot\bm{s}$, and thus by substituting this for $A$ in Eq.~\eqref{eq:formulas_me_A}, we obtain the normal-state part
\begin{align}
\bar{\epsilon}_n(\bm{h})&=\epsilon_n-\bm{h}_n\cdot\bm{\sigma}+O(h^2).
\end{align}
This means that the originally degenerate bands acquire a Zeeman splitting, with the effective magnetic field
\begin{align}
\bm{h}_n&=[\hat{g}\cdot\bm{h}]\hat{g}+n\frac{t_\perp}{R}\hat{g}\times[\bm{h}\times\hat{g}],\notag\\
&\equiv\bm{h}_\parallel+n\frac{t_\perp}{R}\bm{h}_\perp,\label{eq:effective_Zeeman}
\end{align}
where $\partial {h}^i_{n}/\partial h_j$ corresponds to the ($\bm{k}$-dependent) g-factor of the band $n$.
Here, we defined the unit vector $\hat{g}=\bm{g}/g$ and
the component of the magnetic field parallel and perpendicular to the g-vector as 
\begin{align}
\bm{h}_\parallel\equiv[\hat{g}\cdot\bm{h}]\hat{g},\quad \bm{h}_\perp\equiv\hat{g}\times[\bm{h}\times\hat{g}]=\bm{h}-\bm{h}_\parallel,
\end{align}
respectively.
The interband Berry connection for the Zeeman field is given by
\begin{align}
h_i[iA_{\inter}^{h_i}]_{n,-n}&=\frac{ng}{2R^2}\bm{h}_\perp\cdot\bm{\sigma},\label{eq:interbandA_zeeman}
\end{align}
which simply means that the magnetic field perpendicular to the g-vector causes the interband transitions.

To illustrate QGPP for the Zeeman field, 
we consider two examples of the superconducting order parameter that typically appear in bilayer superconductors.
One is the even-parity order parameter
\begin{subequations}\begin{align}
\Delta_{\rm e}=\psi+\bm{d}\cdot\bm{s}\,\eta_z,
\label{eq:eveparity_orderparameter_Zeeman}
\end{align}
and the other is the odd-parity order parameter
\begin{align}
\Delta_{\rm o}=\psi\,\eta_z+\bm{d}\cdot\bm{s}.
\label{eq:oddparity_orderparameter_Zeeman}
\end{align}\label{eq:orderparameters_zeeman}\end{subequations}
The first component of the even-parity order parameter $\Delta_{\rm e}$ is the spin-singlet pair potential $\psi$.
The other component $\bm{d}\cdot\bm{s}\eta_z$
is the spin-triplet pair-density-wave state, which belongs to an even-parity representation and thus is generally admixed.
The first component $\psi\eta_z$ in the odd-parity order parameter $\Delta_{\rm o}$ is the pair-density-wave state, while the second component corresponds to the spin-triplet pairing state.
The realization of the order parameter like $\Delta_{\rm o}$ has been theoretically predicted in bilayer superconductors~\cite{Yoshida2012PDW,Yanase2022-pw} and has recently been discussed for the high-field superconducting phase of a two-phase superconductor CeRh$_2$As$_2$~\cite{Khim2021-er,Fischer2023-rf}.

In GBR with MCBB, the even-parity order parameter is given by
\begin{align}
[\Delta_{\rm e,\B}]_{nn'}&=\delta_{nn'}\left[\psi +\frac{ng}{R}\bm{d}\cdot\hat{g}\right] \notag \\
&\quad+\delta_{n,-n'}\left[-\bm{d}\cdot\hat{g}\frac{t_\perp}{R}-in\bm{d}\times\hat{g}\cdot\bm{\sigma}\right],\label{eq:OP_Zeeman_even}
\end{align}
from Eqs.~\eqref{eq:formulas_me_A} and~\eqref{eq:eveparity_orderparameter_Zeeman}.
The odd-parity order parameter is given by
\begin{align}
[\Delta_{\rm o,\B}]_{nn'}&=\delta_{nn'}\left[\psi\frac{ng}{R}\hat{g}+\bm{d}_\parallel+n\frac{t_\perp}{R}\bm{d}_\perp\right]\cdot\bm{\sigma}\notag\\
&\quad+{\delta_{n,-n'}}\left[-\psi\frac{t_\perp}{R}\hat{g}+\frac{g}{R}\bm{d}_\perp\right]\cdot\bm{\sigma},\label{eq:OP_Zeeman_odd}
\end{align}
from Eqs.~\eqref{eq:formulas_me_A} and~\eqref{eq:oddparity_orderparameter_Zeeman},
by introducing $\bm{d}_\parallel=[\bm{d}\cdot\hat{g}]\hat{g}$ and $\bm{d}_\perp\equiv\hat{g}\times[\bm{d}\times\hat{g}]$.
The intraband components of the even- and odd-parity order parameters are purely pseudospin-singlet and -triplet, respectively, in accordance with the inversion symmetry.

By using Eq.~\eqref{eq:interbandA_zeeman}, the field-linear QGPP is given by
\begin{align}
[\Delta_{\rm e,\geom}(\bm{h})]_{nn'}
&=-[ih_iA^{h_i}_\inter]_{n,-n}[\Delta_{\rm e,\B}]_{-n,n'}\notag\\
&\qquad\quad-[\Delta_{\rm e,\B}]_{n,-n'}[ih_iA^{h_i}_\inter]_{-n',n'}\label{eq:41}\\
&=\delta_{nn'}\left[\frac{-i}{R^2}\bm{h}\cdot\bm{d}\times\bm{g}\right]
-\delta_{n,-n'}\left[\psi\frac{ng}{R^2}\bm{h}_\perp\cdot\bm{\sigma}\right]\notag,
\end{align}
for the even-parity order parameter.
For the odd-parity order parameter we obtain
\begin{align}
[\Delta_{\rm o,\geom}(\bm{h})]_{nn'}
&=\delta_{nn'}\left[-i\psi\frac{ngt_\perp}{R^3}\hat{g}\times\bm{h}+i\frac{ng^2}{R^3}\bm{d}_\perp\times\bm{h}_\perp\right]\cdot\bm{\sigma}\notag\\
&\quad+\delta_{n,-n'}\left[i\frac{g\psi}{R^3}\bm{h}\times\bm{g}+i\frac{gt_\perp }{R^3}\bm{h}_\perp\times\bm{d}_\perp\right]\cdot\bm{\sigma}.
\end{align}

The obtained QGPP indicates that chiral superconductivity is effectively realized for the even-parity order parameter, in a way similar to the Abelian case discussed previously (Sec.~\ref{sec:non-s-wave}).
Actually, the BdG Hamiltonian in GBR$+$MCBB has the form
\begin{subequations}
    \begin{align}
\bar{H}_\B(\bm{h})&=\oplus_{n=\pm}\begin{pmatrix}
    \epsilon_n-\bm{h}_n\cdot\bm{\sigma}&\psi_{n}(\bm{h})\\\psi_n^*(\bm{h})
    &-\epsilon_n-\bm{h}_n\cdot\bm{\sigma}
\end{pmatrix},\\
\psi_n(\bm{h})&\equiv \psi+\frac{ng}{R}\bm{d}\cdot\hat{g}-i\frac{1}{R^2}\bm{h}\cdot\bm{d}\times\bm{g},
\end{align}
\end{subequations}
when the interband order parameter $[\bar{\Delta}_\B(\bm{h})]_{n,-n}$ is neglected as validated for $|\psi|\ll R$.
On the other hand, non-unitary spin-triplet states are effectively realized for the odd-parity pairing.
By writing the BdG Hamiltonian in GBR$+$MCBB as
\begin{align}
\bar{H}_\B(\bm{h})&=\oplus_{n=\pm}\begin{pmatrix}
    \epsilon_n-\bm{h}_n\cdot\bm{\sigma}&\bm{d}_{n}(\bm{h})\cdot\bm{\sigma}\\\bm{d}_n^*(\bm{h})\cdot\bm{\sigma}
    &-\epsilon_n-\bm{h}_n\cdot\bm{\sigma}
\end{pmatrix},
\end{align}
neglecting $[\bar{\Delta}_\B(\bm{h})]_{n,-n}$,
we obtain the effective spin-triplet order parameter, for example,
\begin{align}
\bm{d}_n(\bm{h})=\psi\frac{ng}{R}\left[\hat{g}+i\frac{t_\perp}{R^2}\bm{h}\times\hat{g}\right],
\end{align}
for the purely pair-density-wave state $\psi\neq0$ and $\bm{d}=0$,
and its nonunitarity is represented by
\begin{align}
\bm{d}_{n}(\bm{h})\times\bm{d}_{n}^*(\bm{h})&=-2i|\psi|^2\frac{t_\perp g^2}{R^4}\bm{h}_\perp.\label{eq:mag}
\end{align}
For the case of the staggered Rashba spin-orbit coupling $\bm{g}(\bm{k})\sim (-k_y,k_x,0)$, the perpendicular magnetic field $\bm{h}_\perp=(0,0,h_z)$ gives rise to the 
the perpendicular Cooper-pair magnetization as expected from Eq.~\eqref{eq:mag}.

\subsubsection{Supercurrent}
For the case of the supercurrent, we consider an even-parity order parameter
\begin{align}
\Delta_{\rm e2}=\psi+\psi'\eta_x,
\end{align}
to highlight the difference from the Zeeman field. 
The order parameters $\psi$ and $\psi'$ represent coexisting spin-singlet Cooper pairs within and between the layers, respectively, and the GBR+MCBB expression is given by
\begin{align}
[\Delta_{\rm e2,\B}]_{nn'}=\delta_{nn'}\Bigl[\psi+\frac{nt_\perp}{R}\psi'\Bigr]+\delta_{n,-n'}\Bigl[\frac{g}{R}\psi'\Bigr].\label{eq:Deltae2}
\end{align}

The system preserves combined inversion and time-reversal symmetry ($\Theta I$ symmetry) even under the supercurrent, and therefore the spectrum remains degenerate:
\begin{align}
\bar{\epsilon}_n(\bm{q})=\epsilon_n+\bm{q}\cdot{\bm{v}}_n+O(q^2),
\end{align}
whose leading-order term represents the Doppler shift with the group velocity
$\bm{v}_n\equiv\nabla_{\bm{k}}\epsilon_n.$
The interband Berry connection for the supercurrent coincides with the usual Berry connection and is given by
\begin{align}
q_i[iA^{i}_{\inter}]_{n,-n}&=-\frac{ng[\bm{q}\cdot\nabla_{\bm{k}}]t_\perp}{2R^2}+\frac{nt_\perp[\bm{q}\cdot\nabla_{\bm{k}}]g}{2R^2}\notag\\
&\qquad\qquad+i\frac{[\bm{q}\cdot\nabla_{\bm{k}}\bm{g}]\times\hat{g}}{2R}\cdot\bm{\sigma}.
\end{align}

The field-linear QGPP for the even-parity order parameter $\Delta_{\rm e2}$ in Eq.~\eqref{eq:Deltae2} is given by
\begin{align}
[\Delta_{\rm e2,\geom}(\bm{q})]_{nn'}&=\delta_{nn'}\left[-i\psi'\frac{[\bm{q}\cdot\nabla_{\bm{k}}\bm{g}]\times\bm{g}}{R^2}\cdot\bm{\sigma}\right]\\
&\quad+\delta_{n,-n'}\left[-2\psi q_i[iA^i_{\inter}]_{n,-n}\right]\notag.
\end{align}
The intraband component effectively realizes the so-called anapole superconductivity~\cite{Kitamura2023-on,Kanasugi2022-cl} in accordance with the preserved $\Theta I$ symmetry.
Indeed, the effective intraband order parameter 
\begin{align}
[\bar{\Delta}_{\rm e2,\B}(\bm{q})]_{nn}&=\psi+\frac{nt_\perp}{R}\psi'-i\psi'\frac{[\bm{q}\cdot\nabla_{\bm{k}}\bm{g}]\times\bm{g}}{R^2}\cdot\bm{\sigma},
\end{align}
is a $s+ip$-wave state when $\psi$ and $\psi'$ are $\bm{k}$-independent and $\bm{g}(\bm{k})$ is linear in $\bm{k}$.
This result indicates the engineering of the pseudospin-triplet component, out of only the even-parity spin-singlet Cooper pairs.

In closing the section, we make a comment on the covariant-derivative term of QGPP $-i[A^{X_i}_{\inter;X_j},\Delta_\B]$ in Eq.~\eqref{eq:Delta_g_general}.
For the case of the supercurrent $\bm{X}=\bm{q}$, the covariant derivative of the Berry connection is given by
\begin{align}
[iA^{i}_{\inter;j}]_{n,-n}&=\frac{nt_\perp}{2R^2}\left[\hat{g}\cdot\partial_{k_i}\partial_{k_j}\bm{g}-\frac{2g}{R^2}\partial_{k_i}g\partial_{k_j}g\right]\label{eq:covariantderivative_spinfulbilayer}
\\
&\quad+\left[\frac{ig^3}{2R^3}\partial_{k_i}\partial_{k_j}\hat{g}+\frac{it_\perp^2}{2R^3}\partial_{k_i}\partial_{k_j}\bm{g}\right]\times\hat{g}\cdot\bm{\sigma},\notag
\end{align}
and coincides with the usual definition, i.e., that for the wave-number space.
Here we neglected the $\bm{k}$ derivatives of $t_\perp$ for simplicity.
This expression indicates that the term $-i[A^{i}_{\inter;j},\Delta_\B]$ generally becomes finite
as is explicitly confirmed for the order parameters in GBR given in Eqs.~\eqref{eq:OP_Zeeman_even}, \eqref{eq:OP_Zeeman_odd}, and~\eqref{eq:Deltae2},
leading to a quantum-geometric contribution to equilibrium properties.

\section{Summary}
In this paper, we have elucidated how the quantum geometry of normal electrons is encoded to Cooper pairs, by introducing the notion of quantum-geometric pair potential (QGPP).
{QGPP describes the change of the pair-potential in the band representation in response to the external field, and is essential to understand the superconducting properties in the external field. We}
have clarified that QGPP is solely responsible for the quantum-geometric corrections to the thermodynamic coefficients {for a given pair potential in the orbital basis}.
QGPP is introduced in the generalized band representation (GBR), which ensures an explicit gauge covariance for each Taylor coefficient in terms of the external field.
We illustrated the basic ideas based on the system of nondegenerate bands from plain $s$-wave superconductivity to arbitrary pair potentials and then discussed its extension to the system with degenerate and/or entangled bands.
GBR with QGPP in multiorbital systems offers a concise analytical method to reduce the system's degrees of freedom and derive the effective Hamiltonian, which helps to understand the equilibrium properties of the system based on the known results of the simpler models.
QGPP describes the correction to the pair potential from the external field and quantum geometry, not only contributing to thermodynamic properties but also offering a guiding principle to design exotic superconducting states such as chiral superconductivity, non-unitary spin-triplet superconductivity, anapole superconductivity, and topological superconductivity.
QGPP can play a particularly important role in flat-band superconductors and/or superconductors with a shallow Fermi sea, i.e., those in the BCS-BEC regime, as well as in systems with nearly degenerate bands near the Fermi energy.

\begin{acknowledgments}
We appreciate helpful discussions with Hikaru Watanabe.
This work was supported by JSPS KAKENHI (Grants Nos. JP18H05227, JP18H01178, JP20H05159, JP21K13880, JP21K18145, JP21J14804, 	22H04476, JP22H04933, JP22H01181,  JP22J22520, JP23K17353,
JP24H00007,
JP24K21530), JSPS research fellowship, and WISE Program MEXT.

\end{acknowledgments}


%

\appendix

\section{Quantum-geometric expansion}
\label{app:quantum-geometric_expansion}
In this section, we derive quantum-geometric correction to the free energy based on QGPP.
Hereafter, we keep the arguments of the quantities implicit when they are unnecessary.
Based on the discussion in the main text, we start from
\begin{align}
\Omega&=-\int\frac{d^dk}{(2\pi)^d}\frac{1}{2\beta}\tr\ln\left[1+e^{-\beta H_\B}\right]+\text{const.}
\end{align}
In principle, all we have to do is simply expand $H_\B$ in terms of $\Delta_\geom$, an $O(X)$ quantity.
Here, we take another route, which allows a transparent expansion regarding QGPP.
We rewrite $\Omega$ by using the Green's function
\begin{align}
\Omega&=-\int\frac{d^dk}{(2\pi)^d}\frac{1}{2\beta}\sum_{\omega_n}\tr\ln[G_\B^{-1}(i\omega_n)]+\text{const.},
\end{align}
with $G^{-1}_\B=i\omega_n-H_\B $.
Here and hereafter, the convergence factor is abbreviated, which is unnecessary for our purpose.
By defining the Green's function free from the quantum geometry
\begin{align}
G_\conv^{-1}(i\omega_n)\equiv i\omega_n-\begin{pmatrix}\epsilon(\bm{X})&\Delta_0\\\Delta_0^\dagger&-\epsilon(-\bm{X})
\end{pmatrix},
\end{align}
the Dyson equation reads
\begin{align}
    G_\B(i\omega_n)^{-1}&=G_\conv(i\omega_n)^{-1}-\hat{\Delta}_\geom\notag\\
    &=G_\conv(i\omega_n)^{-1}[1-G_\conv(i\omega_n)\hat{\Delta}_\geom],
\end{align}
where we introduced the matrix
\begin{align}
\hat{\Delta}_\geom&\equiv\begin{pmatrix}0&\Delta_\geom\\\Delta_\geom^\dagger&0\end{pmatrix}.
\end{align}
This allows us to separate the contribution of QGPP,
\begin{align}
\Omega&=\Omega_\conv+\Omega_\geom,
\end{align}
with
\begin{align}
\Omega_\conv&=-\int\frac{d^dk}{(2\pi)^d}\frac{1}{2\beta}\sum_{\omega_n}\tr\ln[G_\conv^{-1}(i\omega_n)]+\text{const.},\label{eq_supp:Omega_conv}\\
\Omega_\geom&=-\int\frac{d^dk}{(2\pi)^d}\frac{1}{2\beta}\sum_{\omega_n}\tr\ln[1-G_\conv(i\omega_n)\hat{\Delta}_\geom].\label{eq_supp:Omega_geom}
\end{align}

\subsubsection{Conventional free energy}
We first discuss the conventional free energy $\Omega_\conv$.
By definition, we obtain
\begin{align}
\Omega_\conv(\bm{X})&=[\Omega(\bm{X})]_{\Delta_\geom\to0}\\
&=-\int\frac{d^dk}{(2\pi)^d}\frac{1}{2\beta}\tr\ln\left[1+e^{-\beta H_\conv}\right]+\text{const.},\notag
\end{align}
with
\begin{align}
H_\conv(\bm{k};\bm{X})&=\begin{pmatrix}\epsilon(\bm{X})&\Delta_0\\\Delta_0^\dagger&-\epsilon(-\bm{X})
\end{pmatrix}.
\end{align}
Since this is a direct sum of single-band superconductivity, textbook calculations are available.
We obtain
\begin{align}
\Omega_\conv(\bm{X})&=\Omega_\conv(0)+\Omega_\para(\bm{X})+\Omega_\dia(\bm{X})\\
&=\Omega(0)+\Omega_\para(\bm{X})+\Omega_\dia(\bm{X}),\notag
\end{align}
with
\begin{align}
&\Omega_\para(\bm{X})=\frac{1}{2}X_aX_b\sum_{n}\int_\BZ\frac{d^dk}{(2\pi)^d}\partial_{X_a}\epsilon_n\partial_{X_b}\epsilon_n\,f'(E_n),\notag\\
&\Omega_\dia(\bm{X})=\frac{1}{4}X_aX_b\sum_n\int_\BZ\frac{d^dk}{(2\pi)^d}\partial_{X_a}\partial_{X_b}\epsilon_n\notag\\
&\qquad\qquad\qquad\qquad\cdot\left[1-\frac{\epsilon_n}{E_n}\tanh\frac{\beta E_n}{2}\right].
\end{align}
We used the fact that $\sum_n\int_\BZ\frac{d^dk}{(2\pi)^d}\partial_{X_i}\partial_{X_j}\epsilon_n$ vanishes, which can be checked explicitly for $X=\bm{q},\bm{h}$. [If this does not hold, the $\frac{1}{2}\int_\BZ\frac{d^dk}{(2\pi)^d}\tr[H_{\rm N}]$ term included in the ``$\text{const.}$" in the free energy depends on $\bm{X}$. 
Taking account of this contribution, the above results are reproduced by the cancellation.
Note also that the time-reversal symmetry is assumed in the absence of $\bm{X}$, leading to vanishing $O(X)$ contribution.]

\subsubsection{Quantum-geometric free energy}
We obtain the expansion of $\Omega_\geom$ as follows:
\begin{subequations}
\begin{align}
\Omega_\geom=\Omega_1+\Omega_2+O(X^3),
\end{align}
with
\begin{align}
\Omega_1&=\int\frac{d^dk}{(2\pi)^d}\frac{1}{2\beta}\sum_{\omega_n}\tr[G_\conv(i\omega_n)\hat{\Delta}_\geom],\\
\Omega_2&=\int\frac{d^dk}{(2\pi)^d}\frac{1}{4\beta}\sum_{\omega_n}\tr[(G_\conv(i\omega_n)\hat{\Delta}_\geom)^2].
\end{align}\label{eq_supp:Omega_g_all}    \end{subequations}
To evaluate the first term, note that $G_{\rm c}(i\omega_n)$ includes only the intraband component.
Accordingly, only the intraband component of $\Delta_\geom$ contributes, which is $O(X^2)$.
Thus, we can replace $G_{\rm c}(i\omega_n)$ with the Green's function in the absence of the field $\bm{X}$.
By writing its $n$-th band component as
\begin{align}
g_n(i\omega_n)^{-1}=i\omega_n-\epsilon_n\tau_z-\Delta_0\tau_x,
\end{align}
with the Pauli matrices in the {Nambu} space $\tau_\mu$,
we obtain
\begin{align}
\!\Omega_1&=\int\frac{d^dk}{(2\pi)^d}\frac{1}{2\beta}\sum_{\omega_n,n}\tr_\tau[\tau_x\,g_n(i\omega_n)][-\Delta_0X_aX_b\QM^{X_aX_b}_n]\notag\\
&=\int\frac{d^dk}{(2\pi)^d}\frac{1}{\beta}\sum_{\omega_n,n}\frac{[-\Delta_0^2X_aX_b\QM^{X_aX_b}_n]}{(i\omega_n-E_n)(i\omega_n+E_n)}\notag\\
&=\frac{1}{2}X_aX_b\int\frac{d^dk}{(2\pi)^d}\sum_n\Delta_0^2\,\QM^{X_aX_b}_n\frac{\tanh\beta E_n/2}{E_n}.\label{eq_supp:Omega_1}
\end{align}
This term makes a dominant contribution to the superfluid weight of the flat-band superconductivity~\cite{Liang2017-ax}.
{Here, we defined the quantum metric $\QM_n^{X_aX_b}\equiv\sum_{m\neq n}A_{nm}^{X_a}A_{mn}^{X_b}+\text{c.c.}$, with the Berry connection $A^{X_a}_{nm}\equiv-i\braket{u_n(\bm{k};\bm{X})|\partial_{X_a}u_m(\bm{k};\bm{X})}|_{X\to0}$.}

To evaluate the second term of $\Omega_\geom$, $G_\conv(i\omega_n)$ can again be replaced with those at $\bm{X}=0$, since $\Delta_\geom=O(X)$.
In particular, only the interband component of $\Delta_\geom$ is $O(X)$, and thus we obtain
\begin{align}
\Omega_2&=\int\frac{d^dk}{(2\pi)^d}\frac{1}{4\beta}\sum_{\omega_n,n\neq m}\notag\\
&\quad\tr_\tau[g_n(i\omega_n)[\Delta_\geom]_{nm}g_m(i\omega_n)[\Delta_\geom]_{mn}]\notag\\
&=\int\frac{d^dk}{(2\pi)^d}\frac{X_aX_b}{4}\sum_{n\neq m}8\Delta_0^2A_{nm}^{X_a}A_{mn}^{X_b}\notag\\
&\qquad\qquad\cdot\frac{1}{\beta}\sum_{\omega_n}\frac{-\Delta_0^2+(i\omega_n+\epsilon_n)(i\omega_n-\epsilon_m)}{(\omega_n^2+E_n^2)(\omega_n^2+E_m^2)}.
\end{align}
To obtain the second line, note that $\omega_n$-odd components vanish.
By using the decomposition such as
\begin{align}
&-\frac{i\omega_n+\epsilon_n}{\omega_n^2+E_n^2}=\frac{1}{2}\sum_{s=\pm1}\left(1+\frac{s\epsilon_n}{E_n}\right)\frac{1}{i\omega_n-sE_n},
\end{align}
we obtain
\begin{align}
&\frac{1}{\beta}\sum_{\omega_n}\frac{-\Delta_0^2+(i\omega_n+\epsilon_n)(i\omega_n-\epsilon_m)}{(\omega_n^2+E_n^2)(\omega_n^2+E_m^2)}\notag\\
&=\frac{1}{4}\sum_{s,s'=\pm1}\frac{ss'\Delta_0^2}{E_nE_m}\frac{f(sE_n)-f(s'E_m)}{sE_n-s'E_m}\\
&\quad+\frac{1}{4}\sum_{s,s'=\pm1}\left(1+\frac{s\epsilon_n}{E_n}\right)\left(1-\frac{s'\epsilon_m}{E_m}\right)\frac{f(sE_n)-f(s'E_m)}{sE_n-s'E_m}.\notag
\end{align}
Since this is symmetric with respect to the permuation of $n$ and $m$, we can replace $X_aX_bA_{nm}^{X_a}A_{mn}^{X_b}$ with
\begin{align}
X_aX_b\,\mathrm{ Re}\left[A_{nm}^{X_a}A_{mn}^{X_b}\right]=\frac{X_aX_b}{2}\QM_{nm}^{X_aX_b},
\end{align}
{with $\QM_{nm}^{X_aX_b}\equiv A_{nm}^{X_a}A_{mn}^{X_b}+\text{c.c.}$.}
We can show $\Omega_2=\Omega_\geom-\Omega_1$ with the expression of $\Omega_\geom$ given in Eq.~\eqref{eq:Omega_all}.
Actually, we can evaluate $\Omega_2$ as follows:
\begin{widetext}
\begin{align*}
\Omega_2&=\int\frac{d^dk}{(2\pi)^d}\frac{X_aX_b}{4}\sum_{n\neq m,s,s'}4\Delta_0^2\QM_{nm}^{X_aX_b}\frac{1}{4}\left[\frac{-ss'\Delta_0^2}{E_nE_m}+\left(1+\frac{s\epsilon_n}{E_n}\right)\left(1-\frac{s'\epsilon_m}{E_m}\right)\right]\frac{f(sE_n)-f(s'E_m)}{sE_n-s'E_m}\\
&=\int\frac{d^dk}{(2\pi)^d}\frac{X_aX_b}{8}\Delta_0^2\sum_{n\neq m,s,s'}\QM_{nm}^{X_aX_b}\left[\frac{-2ss'\Delta_0^2+2(E_n+s\epsilon_n)(E_m-s'\epsilon_m)}{E_nE_m}\right]\frac{f(sE_n)-f(s'E_m)}{sE_n-s'E_m}\\
&=\int\frac{d^dk}{(2\pi)^d}\frac{X_aX_b}{8}\Delta_0^2\sum_{n\neq m,s,s'}\QM_{nm}^{X_aX_b}\left[\frac{-2ss'\Delta_0^2+2E_nE_m-2ss'\epsilon_n\epsilon_m+2(s\epsilon_nE_m-s'\epsilon_mE_n)}{E_nE_m}\right]\frac{f(sE_n)-f(s'E_m)}{sE_n-s'E_m}\\
&=\int\frac{d^dk}{(2\pi)^d}\frac{X_aX_b}{8}\Delta_0^2\sum_{n\neq m,s,s'}\QM_{nm}^{X_aX_b}\left[\frac{-2ss'\Delta_0^2+2E_nE_m-2ss'\epsilon_n\epsilon_m}{E_nE_m}\right]\frac{f(sE_n)-f(s'E_m)}{sE_n-s'E_m}\\
&=\int\frac{d^dk}{(2\pi)^d}\frac{X_aX_b}{8}\Delta_0^2\sum_{n\neq m,s,s'}\QM_{nm}^{X_aX_b}\left[ss'\frac{(\epsilon_n-\epsilon_m)^2-(sE_n-s'E_m)^2}{E_nE_m}\right]\frac{f(sE_n)-f(s'E_m)}{sE_n-s'E_m}\\
&=\Omega_\geom-\int\frac{d^dk}{(2\pi)^d}\frac{X_aX_b}{8}\Delta_0^2\sum_{n\neq m,s,s'}\QM_{nm}^{X_aX_b}\left[\left(\frac{s'}{E_m}-\frac{s}{E_n}\right)[f(sE_n)-f(s'E_m)]\right]\\
&=\Omega_\geom-\int\frac{d^dk}{(2\pi)^d}\frac{X_aX_b}{8}\Delta_0^2\sum_{n\neq m,s,s'}\QM_{nm}^{X_aX_b}\left(-\frac{s'}{E_m}f(s'E_m)-\frac{s}{E_n}f(sE_n)\right)\\
&=\Omega_\geom-\Omega_1.
\end{align*}
\end{widetext}
We used the decomposition
\begin{align}
 ss'(\epsilon_n-\epsilon_m)^2&=-2ss'\Delta_0^2+ss'(sE_n-s'E_m)^2\notag\\
 &\qquad+2E_nE_m-2ss'\epsilon_n\epsilon_m,
\end{align}
and the fact that the term $(s\epsilon_nE_n-s'\epsilon_mE_n)$ is odd in the permutation of $(n,s)\leftrightarrow(m,s')$ and thus does not contribute.
Thus, we reproduce $\Omega_1+\Omega_2=\Omega_{\geom}$ in the main text.

\section{Derivation of QGPP (Abelian case)}
\label{app:Delta_g_derivation}
Here we derive QGPP in the absence of band degeneracy up to the first order for readers' convenience.
We abbreviate the arguments $\bm{k}$ in the following.
Let us define the Wilson line through
\begin{align}
\gamma_n(\bm{X})\equiv\int_0^{\bm{X}} d\bar{X}_j\,A^{X_j}_{nn}(\bar{\bm{X}}),
\end{align}
where the integral is taken along the straight line $0\to\bm{X}$.
Here, the Berry connection is defined by
\begin{align}
A_{nm}^{X_i}(\bm{X})\equiv-i\braket{u_n(\bm{X})|\partial_{X_i}|u_m(\bm{X})},
\end{align}
and we use $A_{nm}^{X_i}\equiv A_{nm}^{X_i}(0)$ in the following.
By using $\theta_{nm}(\bm{X})\equiv\gamma_n(\bm{X})-\gamma_m(-\bm{X})$, we obtain
\begin{align}
&\partial_{X_i}[e^{i\theta_{nm}(\bm{X})}\braket{u_n(\bm{X})|\Delta|u_m(-\bm{X})}]\notag\\
&=e^{i\theta_{nm}(\bm{X})}\Bigl\{i\partial_{X_i}\theta_{nm}(\bm{X})\braket{u_n(\bm{X})|\Delta|u_m(-\bm{X})}\\
&\ +\braket{\partial_{X_i}u_n(\bm{X})|\Delta|u_m(-\bm{X})}+\braket{u_n(\bm{X})|\Delta|\partial_{X_i}u_m(-\bm{X})}\Bigr\}.\notag
\end{align}
We obtain
\begin{align}
\lim_{X\to0}\partial_{X_i}\gamma_n(\bm{X})&=A^{X_i}_{nn},
\end{align}
and thus
\begin{align}
\lim_{X\to0}\partial_{X_i}\theta_{nm}(\bm{X})&=A^{X_i}_{nn}+A^{X_i}_{mm}.
\end{align}
We also obtain
\begin{align}
&\braket{\partial_{X_i}u_n(\bm{X})|\Delta|u_m(\bm{X})}\notag\\
&=\quad\sum_l\braket{\partial_{X_i}u_n(\bm{X})|u_l(\bm{X})}\braket{u_l(\bm{X})|\Delta|u_m(\bm{X})}\notag\\
&=-iA_{nl}^{X_i}(\bm{X})[\Delta_\B(\bm{X})]_{lm},
\end{align}
and so on.
Thus, we obtain
\begin{align}
&\lim_{X\to0}\partial_{X_i}[e^{i\theta_{nm}(\bm{X})}\braket{u_n(\bm{X})|\Delta|u_m(-\bm{X})}]\notag\\
&=-iA^{X_i}_{nl}[\Delta_\B]_{lm}-i[\Delta_\B]_{nl} A_{lm}^{X_i}\notag\\
&\qquad+i(A_{nn}^{X_i}+A_{mm}^{X_i})[\Delta_\B]_{nm}\notag\\
&=-i\{A^{X_i}_\inter,\Delta_\B\}.
\end{align}
Higher-order derivatives can also be calculated by using the formula for the Abelian Wilson line,
\begin{align}
\partial_{X_i}\gamma_n(\bm{X})&=A_{nn}^{X_i}(\bm{X})+\int_0^1dt\,t\,\Omega^{X_iX_j}_{nn}(t\bm{X})X_j,
\label{eq:magnetic_field_formula}
\end{align}
with the Berry curvature $\Omega^{X_iX_j}_{nn}(\bm{X})\equiv\partial_{X_i}A_{nn}^{X_j}(\bm{X})-\partial_{X_j}A_{nn}^{X_i}(\bm{X})$,
and the results corresponding to Eq.~\eqref{eq:Delta_g_general} are obtained.
For derivation of Eq.~\eqref{eq:magnetic_field_formula}, follow the discussion in the Appendix of Ref.~\onlinecite{Luttinger1951-hj} and replace the vector potential with the Berry connection ${A}_{nn}^{X_i}(\bm{X})$.
The derivation for the Non-Abelian version of Eq.~\eqref{eq:magnetic_field_formula} is given in Appendix~\ref{subsec:QGP_proof}.

\subsection{QGPP for the two-band model}
\label{app:QGPP_for_the_two-band_model}
Here we illustrate QGPP for the two-band model
\begin{align}
H_{\rm N}=\xi(\bm{X})+\bm{g}(\bm{X})\cdot\bm{s}=\sum_{n=\pm}\epsilon_n(\bm{X})P_n(\bm{X}),
\end{align}
with $\epsilon_n(\bm{X})=\xi(\bm{X})+n\,g(\bm{X})$ and
\begin{align}
P_n\equiv\frac{1+n\hat{g}(\bm{X})\cdot\bm{s}}{2}.
\end{align}
In particular, we calculate the intraband component of QGPP.
Abbreviating the argument $\bm{X}$, we obtain
\begin{align}
[\partial_{X_i}\Delta_\geom]_{nn}&=-i\sum_{l\neq n}A_{nl}^{X_i}[\Delta_\B]_{ln}+[\Delta_\B]_{nl}A_{ln}^{X_i}\\
&=\sum_{l\neq n}\frac{\braket{u_n|\partial_{X_i}H_{\rm N}|u_l}\braket{u_l|\Delta|u_n}-
(n\leftrightarrow l)}{\epsilon_n-\epsilon_l}\notag\\
&=\frac{1}{2ng}\Bigl\{\tr[P_n\partial_{X_i}H_{\rm N}P_{-n}\Delta]-(n\to-n)\Bigr\}.\notag
\end{align}
After calculating the trace for
\begin{align}
\Delta=\psi+\bm{d}\cdot\bm{s},
\end{align}
we obtain
\begin{align}
[\Delta_\geom]_{nn}&=iX_i\frac{\hat{g}\times\partial_{X_i}\bm{g}\cdot\bm{d}}{g}.
\end{align}
For the case of the Zeeman field $\bm{X}=\bm{h}$, we obtain
\begin{align}
\bm{g}(\bm{X})&=\bm{g}(\bm{h})={\bm{g}-\bm{h}},
\end{align}
and thus $X_i\partial_{X_i}\bm{g}(\bm{X})=-\bm{h}.$
QGPP is given by
\begin{align}
[\Delta_\geom]_{nn}&=-i\frac{\hat{g}\times\bm{h}\cdot\bm{d}}{g}.
\end{align}
In the case of the supercurrent, we obtain
\begin{align}
\bm{g}(\bm{k};\bm{X})&=\bm{g}(\bm{k}+\bm{q}),
\end{align}
and thus $X_i\partial_{X_i}\bm{g}(\bm{X})=\bm{q}\cdot\nabla_{\bm{k}}\bm{g}(\bm{k})$.
QGPP is given by
\begin{align}
[\Delta_\geom]_{nn}&=i\frac{\hat{g}\times[\bm{q}\cdot\nabla_{\bm{k}}]\bm{g}\cdot\bm{d}}{g}.
\end{align}

\section{Derivation of QGPP (general case)}
\label{app:derivation_general}

Here we derive QGPP for the general case in the presence of band degeneracy.
Let us consider $H_{\rm N}$ generally with degeneracy in the absence of the time-reversal breaking field $\bm{X}$.
We denote its eigenstates by $\ket{u_n^{(\lambda)}}$, where the index $\lambda=1,2,\cdots N_n$ distinguishes the eigenstates with the same energy:
$H_{\rm N}\ket{u_n^{(\lambda)}}=\epsilon_n\ket{u_n^{(\lambda)}}$.
They can generally have different energies under the field $\bm{X}$,
\begin{align}
H_{\rm N}(\bm{X})\ket{u_n^{(\lambda)}(\bm{X})}=\epsilon_n^{(\lambda)}(\bm{X})\ket{u_n^{(\lambda)}(\bm{X})},
\end{align}
with $\epsilon^{(\lambda)}_n(\bm{X}=0)=\epsilon_n$.
Let us define
\begin{align}
U_{\bm{X}}^\dagger=\left(\ket{u_1^\one(\bm{X})},\cdots,\ket{u_1^{(N_1)}(\bm{X})},\ket{u_2^{(1)}(\bm{X})},\cdots\right).
\end{align}
For the latter convenience, we allow an additional arbitrary unitary transformation within each originally degenerate space.
Thus, the redefined $\ket{u_n^{(\lambda)}(\bm{X})}$ may not be an exact eigenstate of $H_{\rm N}(\bm{X})$, while $H_{\rm N}(\bm{X})\ket{u_n^{(\lambda)}(\bm{X})}$ belongs to the space spanned by originally degenerate states, i.e., $\{\ket{u_n^{(1)}(\bm{X})},\cdots, \ket{u_n^{(N_n)}(\bm{X})}\}$.
The unitary transformation $U_{\bm{X}}^\dagger$ is also redefined in this way.
Thus, the Hamiltonian in this basis, $U_{\bm{X}}H_{\rm N}(\bm{X})U_{\bm{X}}^\dagger=\epsilon(\bm{X})$, 
is block-diagonal in the originally degenerate space.

By using $U_{\bm{X}}^\dagger$, the Berry connection is given by
\begin{align}
iA^{X_i}(\bm{X})&=U_{\bm{X}}\partial_{X_i}U_{\bm{X}}^\dagger.
\end{align}
We also define $A^{X_i}\equiv\lim_{X\to0}A^{X_i}(\bm{X})$.
The order parameter in the band representation is given by
\begin{align}
\Delta_\B(\bm{X})&=U_{\bm{X}}\Delta U_{-\bm{X}}^\dagger,
\end{align}
since $U_\B(\bm{X})\equiv\diag(U_{\bm{X}},U_{-\bm{X}})$ diagonalizes the normal-state part of the BdG Hamiltonian 
\begin{subequations}\begin{align}
H_\BdG(\bm{X})&=\begin{pmatrix} H_{\rm N}(\bm{X})&\Delta\\\Delta^\dagger&-H_{\rm N}(-\bm{X})\end{pmatrix},\\
H_\B(\bm{X})&\equiv U_\B(\bm{X})H_\BdG(\bm{X}) U_\B(\bm{X})^\dagger\notag\\
&=\begin{pmatrix}
    \epsilon(\bm{X})&\Delta_\B(\bm{X})\\\Delta_\B^\dagger(\bm{X})&-\epsilon(-\bm{X})
\end{pmatrix}.
\end{align}\end{subequations}
The gauge transform of the Bloch states is represented by
\begin{align}
U_{\bm{X}}\to V_{\bm{X}}U_{\bm{X}}.
\end{align}
Here, $V_{\bm{X}}$ is a block-diagonal unitary matrix, which has finite components only between the states originally with the same energy.
The gauge transform here means that we allow the mixture of the states with energies different by $O(X)$.

\subsection{QGPP in GBR}
The order parameter transforms according to $\Delta_\B(\bm{X})\to V_{\bm{X}}\Delta_\B(\bm{X})V_{-\bm{X}}^\dagger$.
On the other hand, we want the QGPP expansion of the form $\Delta_\B(\bm{X})\sim \Delta_\B(0)+O(X)$.
Since $\Delta_\B(0)$ transforms by $V_0$ instead of $V_{\pm\bm{X}}$, we should first construct a quantitiy similar to $\Delta_\B(\bm{X})$ but transforming by $V_0$.
For this purpose, we introduce the Non-Abelian Wilson line operator
\begin{align}
\mathcal{U}_{\bm{X}}&=P\exp\left(-i\int_0^{\bm{X}}d\bar{\bm{X}}\cdot\,A_{\rm{intra}}^{\bar{\bm{X}}}(\bar{\bm{X}})\right)\notag\\
&\equiv \lim_{N\to\infty} W(\bm{X}_N,\bm{X}_{N-1})W(\bm{X}_{N-1},\bm{X}_{N-2})\notag\\
&\qquad\qquad\qquad\cdots W(\bm{X}_2,\bm{X}_1)W(\bm{X}_1,\bm{X}_0),
\end{align}
which is a generalization of $e^{-i\theta_{nm}(\bm{X})}$.
Here, we choose the path to be the straight line $\bm{X}_i=(i/N)\bm{X}$ and defined
\begin{align}
[W(\bm{X}_{i},\bm{X}_j)]_{n\lambda,n'\lambda'}\equiv \braket{u_{n}^{(\lambda)}(\bm{X}_i)| u_{n}^{(\lambda')}(\bm{X}_j)}\delta_{nn'}.
\end{align}
We also defined the Berry connection within the degenerate space,
\begin{subequations}\begin{align}
[A_{\rm{intra}}^{X_i}(\bm{X})]_{nm}&=\delta_{nm}A_{nm}^{X_i}(\bm{X}),\\
A^{X_i}_\intra&\equiv\lim_{X\to0}A^{X_i}_\intra(\bm{X}).
\end{align}\end{subequations}
We introduce the Berry connection connecting the different degenerate spaces by
\begin{subequations}\begin{align}
A^{X_i}_\inter(\bm{X})&\equiv A^{X_i}(\bm{X})-A^{X_i}_{\rm{intra}}(\bm{X}),\\
A^{X_i}_\inter&\equiv\lim_{X\to0}A^{X_i}_\inter(\bm{X}),
\end{align}\end{subequations}
for the latter use.
The Wilson line transforms by
\begin{align}
\mathcal{U}_{\bm{X}}\to V_{\bm{X}}\mathcal{U}_{\bm{X}}V_0^\dagger,
\end{align}
and therefore the quantity
\begin{align}
\bar{\Delta}_\B(\bm{X})&\equiv\mathcal{U}_{\bm{X}}^\dagger \Delta_\B(\bm{X})\mathcal{U}_{-\bm{X}}\notag\\
&=\mathcal{U}_{\bm{X}}^\dagger U_{\bm{X}}\Delta U_{-\bm{X}}\mathcal{U}_{-\bm{X}}^\dagger\notag
\end{align}
transforms by
\begin{align}
\bar{\Delta}_\B(\bm{X})\to V_0\bar{\Delta}_\B(\bm{X})V_0^\dagger,
\end{align}
as desired.
This ensures that each Taylor coefficient of $\bar{\Delta}_\B(\bm{X})$ transforms by $V_0$ and thus is gauge covariant.
In the following, we write
\begin{align}
\Bar{\Delta}_\B(\bm{X})=\bar{U}_{\bm{X}}\Delta\bar{U}_{-\bm{X}}^\dagger,\label{eq:deff_bar_Delta}
\end{align}
by introducing
\begin{align}
\bar{U}_{\bm{X}}=\mathcal{U}_{\bm{X}}^\dagger U_{\bm{X}}.
\end{align}

\subsection{Expansion of QGPP}
Let us expand $\bar{\Delta}_\B(\bm{X})$. At zero-th order, it coincides with $\Delta_\B(0)\equiv\Delta_\B$.
We obtain
\begin{align}
\partial_{X_i}\bar{\Delta}_\B(\bm{X})&=[\partial_{X_i}\bar{U}_{\bm{X}}\bar{U}_{\bm{X}}^\dagger]\bar{U}_{\bm{X}}\Delta \bar{U}_{-\bm{X}}^\dagger\notag\\
&\qquad+\bar{U}_{\bm{X}}\Delta \bar{U}_{-\bm{X}}^\dagger[\bar{U}_{-\bm{X}}\partial_{X_i}\bar{U}_{-\bm{X}}^\dagger]\\
&=-[Q_{X_i}(\bm{X})\bar{\Delta}_\B({\bm{X}})+\bar{\Delta}_\B({\bm{X}})Q_{-X_i}(\bm{X})],\notag
\end{align}
where we introduced the quantum-geometry factor
\begin{align}
Q_{X_i}(\bm{X})\equiv \bar{U}_{\bm{X}}\partial_{X_i}\bar{U}_{\bm{X}}^\dagger,\quad Q_{X_i}\equiv\lim_{X\to0}Q_{X_i}(\bm{X}).
\end{align}
The second-order derivative is given by
\begin{align}
&\partial_{X_i}\partial_{X_j}\bar{\Delta}_\B(\bm{X})\notag\\
&=-[[\partial_{X_j}Q_{X_i}(\bm{X})]\bar{\Delta}_\B({\bm{X}})-\bar{\Delta}_\B({\bm{X}})\partial_{-X_j}Q_{X_i}(-\bm{X})]\notag\\
&\quad-[Q_{X_i}(\bm{X})\partial_{X_j}\bar{\Delta}_\B({\bm{X}})+\partial_{X_j}\bar{\Delta}_\B({\bm{X}})Q_{X_i}(-\bm{X})]\notag\\
&\to-[\partial_{X_j}Q_{X_i},\Delta_\B]+\{Q_{X_i},\{Q_{X_j},\Delta_\B\}\},
\end{align}
for $X\to0$ with $\partial_{X_j}Q_{X_i}\equiv\lim_{X\to0}\partial_{X_j}Q_{X_i}(\bm{X})$.

Let us evaluate the quantum-geometry factor and its derivative.
It can be written as
\begin{align}
Q_{X_i}(\bm{X})&=\mathcal{U}_{\bm{X}}^\dagger U_{\bm{X}}\partial_{X_i}[U_{\bm{X}}^\dagger\mathcal{U}_{\bm{X}}]\notag\\
&=i\,\mathcal{U}_{\bm{X}}^\dagger[A^{X_i}(\bm{X})-\mathcal{A}^{X_i}(\bm{X})]\mathcal{U}_{\bm{X}}.
\end{align}
Here, we 
defined a quantity related to Berry connection
\begin{subequations}
    \begin{align}
i\mathcal{A}^{X_i}(\bm{X})&=\mathcal{U}_{\bm{X}}\partial_{X_i}\mathcal{U}_{\bm{X}}^\dagger,\\
\mathcal{A}^{X_i}&\equiv\lim_{X\to0}\mathcal{A}^{X_i}(\bm{X})=A^{X_i}_\intra.
\end{align}
\end{subequations}
We obtain
\begin{align}
Q_{X_i}=i[A^i-\mathcal{A}^i]=i[A^i-A_{\rm{intra}}^i]= iA_{\rm{inter}}^i.
\end{align}
We also obtain
\begin{align}
\!\!\!\!\partial_{X_j}Q_{X_i}(\bm{X})&=i\mathcal{U}_{\bm{X}}^\dagger i\mathcal{A}^{X_j}(\bm{X})[A^{X_i}(\bm{X})-\mathcal{A}^{X_i}(\bm{X})]\mathcal{U}_{\bm{X}}\notag\\
&-i\mathcal{U}_{\bm{X}}^\dagger [A^{X_i}(\bm{X})-\mathcal{A}^{X_i}(\bm{X})]i\mathcal{A}^{X_j}(\bm{X})\mathcal{U}_{\bm{X}}\notag\\
&+i\mathcal{U}_{\bm{X}}^\dagger \partial_{X_j}[A^{X_i}(\bm{X})-\mathcal{A}^{X_i}(\bm{X})]\mathcal{U}_{\bm{X}}\\
&\to i\Bigl(\partial_{j}[A^{X_i}-\mathcal{A}^{X_i}]+i[\mathcal{A}^{X_i},(A^{X_i}-\mathcal{A}^{X_i})]\Bigr).\notag
\end{align}
The symmetric part with respect to $i,j$ is important for QGPP since we are interested in the combination $\partial_{X_j}Q_{X_i}X_iX_j/2$.
The symmetric part of $\partial_{X_j}Q_i$ is given by
\begin{align}
\frac{\partial_jQ_i+\partial_iQ_j}{2}=\frac{A^{X_i}_{{\rm inter};X_j}+A^{X_j}_{{\rm inter};X_i}}{2},
\end{align}
while $\partial_{X_j}\mathcal{A}^{X_i}$ has an antisymmetric component related to the Berry curvature [See Sec.~\ref{subsec:QGP_proof}].
Here, we defined the covariant derivative of the interband Berry connection
\begin{align}
A^{X_i}_{{\rm inter};X_j}(\bm{X})&\equiv \partial_{X_j}A_{{\rm inter}}^{X_i}(\bm{X})\notag\\
&\qquad\quad+i[A_{{\rm intra}}^{X_j}(\bm{X}),A_{{\rm inter}}^{X_i}(\bm{X})],
\end{align}
and $A^{X_i}_{{\rm inter};X_j}\equiv \lim_{X\to0}A^{X_i}_{{\rm inter};X_j}(\bm{X})$.
Thus, we obtain QGPP by means of the interband Berry connection and its covariant derivative,
\begin{align}
\bar{\Delta}_\B(\bm{X})&=\Delta_\B+\Delta_\geom(\bm{X}),
\end{align}
with
\begin{align}
\Delta_\geom(\bm{X})&=-iX_i\{A^{X_i}_{{\rm inter}},\Delta_\B\}\notag\\
&\qquad+\frac{1}{2}X_iX_j\Bigl(-i[A^{X_i,}_{{\rm inter};X_j},\Delta_\B]\label{eq_supp:Delta_g}\\
&\qquad\qquad-\{A^{X_i}_{{\rm inter}},\{A^{X_j}_{{\rm inter}},\Delta_\B\}\}\Bigr)+O(X^3).\notag
\end{align}

\subsection{GBR and normal-state Hamiltonian}
After the unitary transformation to obtain QGPP, we obtain
\begin{align}
\bar{H}_\B(\bm{X})
&=\begin{pmatrix}\mathcal{U}_{\bm{X}}^\dagger&0\\0&\mathcal{U}_{-\bm{X}}^\dagger\end{pmatrix}H_\B(\bm{X})\begin{pmatrix}\mathcal{U}_{\bm{X}}&0\\0&\mathcal{U}_{-\bm{X}}\end{pmatrix}\notag\\
&=
\begin{pmatrix}\bar{\epsilon}(\bm{X})&\Delta_\B+\Delta_\geom(\bm{X})\\
\Delta_\B^\dagger+\Delta_\geom(\bm{X})^\dagger&-\bar{\epsilon}(-\bm{X})
\end{pmatrix},
\end{align}
with $\bar{\epsilon}(\bm{X})\equiv \bar{U}_{\bm{X}}H_{\rm N}(\bm{X})\bar{U}^\dagger_{\bm{X}}$.
We call this basis the generalized band representation (GBR).
Let us consider the normal-state part
$\bar{\epsilon}(\bm{X})\equiv \mathcal{U}_{\bm{X}}^\dagger\epsilon(\bm{X})\mathcal{U}_{\bm{X}},$
which transforms by $\bar{\epsilon}(\bm{X})\to V_0\bar{\epsilon}(\bm{X})V_0^\dagger$.
We obtain
\begin{align}
\partial_{X_i}\bar{\epsilon}(\bm{X})&=\bar{U}_{\bm{X}}\partial_{X_i}H_{\rm N}(\bm{X})\bar{U}_{\bm{X}}^\dagger-[{Q}_{X_i},\bar{\epsilon}(\bm{X})],\label{eq:D27}
\end{align}
and
\begin{align}
\partial_{X_j}\partial_{X_i}\bar{\epsilon}(\bm{X})&=\bar{U}_{\bm{X}}\partial_{X_i}\partial_{X_j}H_{\rm N}(\bm{X})\bar{U}_{\bm{X}}^\dagger\notag\\
&\quad -[Q_{X_j},\bar{U}_{\bm{X}}\partial_{X_i}H_{\rm N}(\bm{X})U_{\bm{X}}^\dagger]\notag\\
&\quad -[\partial_{X_j}Q_{X_i},\bar{\epsilon}(\bm{X})]-[Q_{X_i},\partial_{X_j}\bar{\epsilon}(\bm{X})]\notag\\
&=\bar{U}_{\bm{X}}\partial_{X_i}\partial_{X_j}H_{\rm N}(\bm{X})\bar{U}_{\bm{X}}^\dagger\label{eq:D28}\\
&\quad -[Q_{X_j},\partial_{X_i}\bar{\epsilon}(\bm{X})+[Q_{X_i},\bar{\epsilon}(\bm{X})]]\notag\\
&\quad -[\partial_{X_j}Q_{X_i},\bar{\epsilon}(\bm{X})]-[Q_{X_i},\partial_{X_j}\bar{\epsilon}(\bm{X})].\notag
\end{align}
We thus obtain, by using $\bar{U}_0=U_0$ and symmetrization of $i$ and $j$,
\begin{align}
\bar{\epsilon}(\bm{X})&=\epsilon+X_i\Bigl(U_0\partial_{X_i}H_{\rm N}U_0^\dagger-[iA_\inter^{X_i},\epsilon]\Bigr)\notag\\
&\quad+\frac{1}{2}X_iX_j\Bigl(U_0\partial_{X_i}\partial_{X_j}H_{\rm N}U_0^\dagger\notag\\
&\quad-[iA_\inter^{X_j},[iA_\inter^{X_i},\epsilon]]-2[iA_\inter^{X_i},\partial_{X_j}\bar{\epsilon}]\notag\\
&\quad -[iA_{\inter;X_j}^{X_i},\epsilon]\Bigr)\notag\\
&=\epsilon+X_i\Bigl(U_0\partial_{X_i}H_{\rm N}U_0^\dagger-[iA_\inter^{X_i},\epsilon]\Bigr)\label{eq:energy_SM}\\
&\quad+\frac{1}{2}X_iX_j\Bigl(U_0\partial_{X_i}\partial_{X_j}H_{\rm N}U_0^\dagger-[iA_{\inter;X_j}^{X_i},\epsilon]\notag\\
&\quad+[iA_\inter^{X_i},[iA_\inter^{X_j},\epsilon]]-2[iA_\inter^{X_i},U_0\partial_{X_j}H_{\rm N}U_0^\dagger]\Bigr),\notag
\end{align}
up to $O(X^2)$.
The obtained Taylor expansion is explicitly gauge-covariant and thus we can arbitrarily choose the eigenstates of $H_{\rm N}$ for $X=0$ to evaluate $\bar{\epsilon}(\bm{X})$ and $\bar{\Delta}_\B(\bm{X})$.

Note that we have not used the fact that $\epsilon$ is proportional to the identity within each space specified by $n$.
Therefore, all the results discussed above remain valid for entangled bands, where the band degeneracy, if any, exists at some discrete points in the Brillouin zone,  including completely degenerate bands as a special case.
In this case, $\epsilon$ is a block-diagonal matrix as is $\bar{\epsilon}(\bm{X})$,
$[\epsilon]_{n\lambda,n'\lambda'}=[\epsilon_n]_{\lambda\lambda'}\delta_{nn'}$.

Equations~\eqref{eq:D27} and~\eqref{eq:D28} give the formulas for calculating the interband Berry connection and its covariant derivative.
Since $\bar{\epsilon}(\bm{X})$ is block-diagonal by construction and does not have matrix elements between different indices $n$, the $(n,n')$ block of the left-hand side of the equations vanishes for $n\neq n'$.
By using Eq.~\eqref{eq:QGF_derivative} for $\partial_{X_j}Q_{X_i}$ derived in the next section and $\Omega_{nn'}^{X_iX_j}=0$ for $n\neq n'$, we obtain the well-known formula
\begin{align}
\!\!\!\![iA^{X_i}_{\inter}]_{nn'}\epsilon_{n'}-\epsilon_n[iA^{X_i}_\inter]_{nn'}&=[U_0\partial_{X_i}H_{\rm N}U_0^\dagger]_{nn'},
\end{align}
and similar one for the covariant derivative,
\begin{align}
&\left[iA^{X_i}_{\inter;X_j}\right]_{n,n'}\!\!\!\!\epsilon_{n'}-\epsilon_n\left[iA^{X_i}_{\inter;X_j}\right]_{n,n'}\notag\\
&=\Bigl(U_0\partial_{X_i}\partial_{X_j}H_{\rm N}U_0^\dagger+[iA_\inter^{X_i},[iA_\inter^{X_j},\epsilon]]\label{eq:CD_formula}\\
&\quad-[iA_\inter^{X_i},U_0\partial_{X_j}H_{\rm N}U_0^\dagger]-[iA_\inter^{X_j},U_0\partial_{X_i}H_{\rm N}U_0^\dagger]\Bigr)_{n,n'},\notag
\end{align}
by noting that the Berry curvature has only the $n=n'$ component.
The formula~\eqref{eq:CD_formula} reproduces that for the Abelian case~\cite{Cook2017-hh},
by using the fact that the $n\neq n'$ component of $[U_0\partial_{X_i}H_{\rm N}U_0^\dagger]_{nn'}$ coincides with $[iA^{X_i}_\inter,\epsilon]_{nn'}$ while its $n=n'$ component corresponds to the group velocity.

\subsection{QGPP by time-reversal symmetric perturbations}
\label{app:QGPP_by_TRSP}
So far, we have considered QGPP induced by time-reversal-odd perturbations.
We can also evaluate QGPP induced by time-reversal-even perturbation $\bm{Y}$.
Note that the normal-state part has the same expression except for $\bm{X}\to\bm{Y}$, since we did not use the time-reversal parity of $\bm{X}$ during the derivation.
Thus, the expansion of the normal-state part is simply obtained by $\bm{X}\to\bm{Y}$ in the formulas.
On the other hand, the order parameter in GBR is given by
\begin{align}
\bar{\Delta}_\B(\bm{Y})&=\bar{U}_{\bm{Y}}\Delta\bar{U}_{\bm{Y}}^\dagger,
\end{align}
where $\bar{U}_{-\bm{X}}^\dagger$ on the right hand side of Eq.~\eqref{eq:deff_bar_Delta} is replaced with $\bar{U}_{\bm{Y}}^\dagger$.
Since this has the same matrix structure as $\bar{\epsilon}(\bm{X})=\bar{U}_{\bm{X}}H_{\rm N}(\bm{X})\bar{U}_{\bm{X}}^\dagger$, we immediately obtain from Eq.~\eqref{eq:energy_SM},
\begin{align}
\bar{\Delta}_\B(\bm{Y})&=\Delta_\B-Y_i[iA^{Y_i}_\inter,\Delta_\B]+\frac{1}{2}Y_iY_j\Bigl(-[iA^{Y_i}_{\inter;Y_j},\Delta_\B]\notag\\
&\qquad\qquad\qquad+[iA^{Y_i}_\inter,[iA^{Y_j}_\inter,\Delta_\B]]\Bigr).
\label{eq:QGPP_for_Y}
\end{align}
In the presence of both $\bm{X}$ and $\bm{Y}$, we can first expand quantities by $\bm{X}$ based on the formulas in the previous sections and then expand them by $\bm{Y}$ by replacing $\Delta_\B\to \bar{\Delta}_\B(\bm{Y})$ and so on.

\subsection{Formula for the quantum-geometry factor}\label{subsec:QGP_proof}
In this section, we show the formula
\begin{align}
&\mathcal{A}^{X_a}(\bm{X})-A^{X_a}_\intra(\bm{X})\label{eq:Stokes_formula}\\
&={\mathcal{U}_{\bm{X}}\left[\int_0^1dt\ t\ \mathcal{U}^\dagger_{{t\bm{X}}}\Omega^{X_aX_b}(t\bm{X})\mathcal{U}_{t\bm{X}}\,X_b\right]\mathcal{U}^\dagger_{\bm{X}}},\notag
\end{align}
following the derivation of the Non-Abelian Stokes' theorem~\cite{Halpern1979-xi,Broda2000-ve}.
Here, we defined the Berry curvature
\begin{align}
\Omega^{X_aX_b}(\bm{X})&=\partial_{X_a}A^{X_b}_\intra(\bm{X})-\partial_
{X_b}A^{X_a}_\intra(\bm{X})\notag\\
&\qquad+i[A^{X_a}_\intra(\bm{X}),A^{X_b}_\intra(\bm{X})].
\end{align}
The relation leads to the formula for the quantum-geometry factor
\begin{align}
Q_{X_a}(\bm{X})&=\mathcal{U}_{\bm{X}}^\dagger \,iA_\inter^{X_a}(\bm{X})\mathcal{U}_{\bm{X}}\\
&\quad-i\int_0^1dt\ t\,\mathcal{U}^\dagger_{t\bm{X}}\Omega^{X_aX_b}(t\bm{X})\mathcal{U}_{t\bm{X}}\,X_b.\notag
\end{align}
As a corollary, it follows that the symmetric part of $\partial_{X_j}Q_{X_i}$ is given by that of $A^{X_i}_{\inter;X_j}$ as used in the previous section:
\begin{align}
\partial_{X_j}Q_{X_i}&=iA^{X_i}_{\inter;X_j}-\frac{i}{2}\Omega^{X_iX_j}.\label{eq:QGF_derivative}
\end{align}

Let us define
\begin{align}
\mathcal{U}(\bm{X}_2,\bm{X}_1)\equiv P\exp\left(-i\int^{\bm{X}_2}_{\bm{X}_1}dX_a'\,A^{X_a}_\intra(\bm{X}')\right),
\end{align}
for convenience, where the path is the straight line between $\bm{X}_1$ and $\bm{X}_2$.
By using this, we can write $\mathcal{U}_{\bm{X}}=\mathcal{U}(\bm{X},0)$ and {$\mathcal{U}(\bm{X}_2,\bm{X}_1)^\dagger=\mathcal{U}(\bm{X}_1,\bm{X}_2)$.} 
Thus, we obtain
\begin{align}
i\mathcal{A}^{X_a}(\bm{X})&=\lim_{\epsilon\to0}\epsilon^{-1}\mathcal{U}_{\bm{X}}\left[\mathcal{U}^\dagger_{\bm{X}+\epsilon\hat{a}}-\mathcal{U}^\dagger_{\bm{X}}\right]\notag\\
&=\lim_{\epsilon\to0}\epsilon^{-1}\left[\mathcal{U}(\bm{X},0)\mathcal{U}(0,\bm{X}+\bm{\epsilon})-1\right],
\end{align}
with $\bm{\epsilon}\equiv \epsilon\hat{a}$ and the unit vector $\hat{a}$ in the $a$ direction.
The first term is given by
\begin{align}
\mathcal{U}(\bm{X},0)\mathcal{U}(0,\bm{X}+\bm{\epsilon})&= \Phi(\bm{X},\bm{\epsilon})\mathcal{U}(\bm{X}+\bm{\epsilon},\bm{X})^\dagger\\
&=\bm{\epsilon}\cdot iA^{\bm{X}}_{\intra}(\bm{X})+\Phi(\bm{X},\bm{\epsilon})+O(\epsilon^2),\notag
\end{align}
with
\begin{align}
\Phi(\bm{X},\bm{\epsilon})&\equiv\mathcal{U}(\bm{X},0)\mathcal{U}(0,\bm{X}+\bm{\epsilon})\mathcal{U}(\bm{X}+\bm{\epsilon},\bm{X})\notag\\
&=\mathcal{U}_{\bm{X}}\,P\exp\left(-i\oint_C dX_a'\,A^{X_a}_\intra(\bm{X}')\right)\mathcal{U}_{\bm{X}}^\dagger\notag\\
&=1+O(\epsilon).
\end{align}
Here, the loop $C$ is the infinitesimal triangle 
\begin{align}
C:\ 0\to\bm{X}\to\bm{X}+\bm{\epsilon}\to0.
\end{align}
In the following, we evaluate the $O(\epsilon)$ contribution in the loop integral $I[C]\equiv P\exp\left(-i\oint_CdX_a'\,A^{X_a}_\intra(\bm{X}')\right)$.

Note that the loop $C$ can be deformed to the following $C'$ without changing the value of $I[C]$.
Let us take $N$ points on the line $0\to\bm{X}$ by $A_i\equiv t_i\bm{X}$ with $t_i=i/N$ $(i=1,\cdots N)$.
We also take $N$ points $B_i\equiv t_i(\bm{X}+\bm{\epsilon})$ on the line $0\to\bm{X}+\bm{\epsilon}$.
The path $C'$ is given by a combination of loops,
\begin{align}
C':\ D_0\to D_1\to D_2\to\cdots \to D_{N-1},
\end{align}
with 
\begin{align}
D_0&:\ 0\to A_1\to B_1\to0,\\
D_i&:\ 0\to B_{i+1}\to B_i\to A_i\to A_{i+1}\to B_{i+1}\to 0,\notag
\end{align}
for $i=1,\cdots N-1$.
Accordingly, we obtain
\begin{align}
I[C]=I[C']=I[D_{N-1}]\cdots I[D_1]I[D_0].
\end{align}
The $(i-1)$-th loop contributes by
\begin{align}
I[D_{i-1}]&=\mathcal{U}(0,B_{i})\mathcal{U}(B_{i},A_{i})\mathcal{U}(A_{i},A_{i-1})\notag\\
&\quad\cdot\mathcal{U}(A_{i-1},B_{i-1})\mathcal{U}(B_{i-1},B_{i})\mathcal{U}(B_{i},0)\\
&=1+i\frac{t_i\epsilon}{N}\mathcal{U}^\dagger_{t_i\bm{X}}\Omega^{X_aX_b}(t_i\bm{X})\mathcal{U}_{t_i\bm{X}}X_b,\notag
\end{align}
neglecting $O(1/N^2,\epsilon^2)$ terms.
Thus, we obtain by $N\to\infty$
\begin{align}
I[C]&=1+i\epsilon\int_0^1 dt\ t\ \mathcal{U}^\dagger_{t\bm{X}}\Omega^{X_aX_b}(t\bm{X})\mathcal{U}_{t\bm{X}}X_b,
\end{align}
up to $O(\epsilon)$ and reproduce Eq.~\eqref{eq:Stokes_formula}.

\section{MCBB for the spin-orbit-coupled bilayer}
\label{app:MCBB}
To evaluate $\bar{\epsilon}(\bm{k};\bm{X})$ and $\bar{\Delta}_\B(\bm{k};\bm{X})$ for a given model Hamiltonian,
we must calculate matrix elements of the form
\begin{align}
[U_0AU_0^\dagger]_{n\sigma,n'\sigma'}=    \braket{u_{n}^{(\sigma)}(\bm{k})|A|u_{n'}^{(\sigma')}(\bm{k})},
\end{align}
which needs a gauge fixing.
For this purpose, it is convenient to choose the MCBB gauge, whose gauge constraint is 
\begin{subequations}\label{eq:gc_all}
\begin{align}    &\braket{\bm{k}=0,+,s|u_{n}^{(1)}(\bm{k})}=a_{\bm{k}}\delta_{s\uparrow}\quad (a_{\bm{k}}>0),\label{eq:gauge_constraint}\\
    &\ket{u_{n}^{(2)}(\bm{k})}=\Theta I\ket{u_{n}^{(1)}(\bm{k})}.\label{eq:gc2}
\end{align}    
\end{subequations}
Here, we defined the orbital
\begin{align}
        \ket{\bm{k}=0,+,s}&=\ket{+}_\eta\otimes\begin{pmatrix}\delta_{s,\uparrow}\\\delta_{s,\downarrow}\end{pmatrix}_s,
\end{align}
where $\ket{+}_\eta=(1,1)_\eta/\sqrt{2}$ is the positive eigenstate of $\eta_x$
for the centrosymmetric bilayer.
In the other models, the $\ket{+}_\eta$ state should be replaced with a local orbital belonging to a one-dimensional representation of the point group.
The gauge constraint~\eqref{eq:gc_all} coincides with the original one~\cite{Fu2015-xu}, which is given by real-space matrix elements, owing to the translational invariance of $\ket{u_{n}^{(1)}(\bm{k})}$.
Importantly, $\ket{\bm{k}=0,+,s}$ transforms in the same way as the spin state $\ket{s_z=s}$ under symmetry operations.
This transformation property is passed down to $\ket{u_{n}^{(\sigma)}(\bm{k})}$, and the pseudospin $\bm{\sigma}$ in the MCBB gauge can be regarded as something close to the real spin $\bm{s}$~\cite{Fu2015-xu}.

Let us introduce $P_{+s}=P_+^\eta\frac{1+s\,s_z}{2}$ with $P_+^\eta\equiv \ket{+}_\eta\bra{+}_\eta$.
The gauge condition Eq.~\eqref{eq:gauge_constraint} can be rewritten as
\begin{subequations}\begin{align}
&P_{+\uparrow}\ket{u_n^{(1)}(\bm{k})}=a_{\bm{k}}\ket{\bm{k}=0,+,\uparrow},\\
&P_{+\downarrow}\ket{u_n^{(1)}(\bm{k})}=0.\label{eq:second_temp}
\end{align}
\end{subequations}
Let us define $P_n^{(\sigma)}(\bm{k})=\ket{u_n^{(\sigma)}(\bm{k})}\bra{u_n^{(\sigma)}(\bm{k})}$ and $P_n(\bm{k})=P_n^{(1)}(\bm{k})+P_n^{(2)}(\bm{k})$.
By the action of $\Theta I$ to Eq.~\eqref{eq:second_temp}, we obtain $P_{+\uparrow}\ket{u_n^{(2)}(\bm{k})}=0$, and thus
\begin{align}
P_{+\uparrow}P_n^{(2)}(\bm{k})=0.
\end{align}
We also obtain $P_n(\bm{k})P_{+\uparrow}P_n^{(1)}(\bm{k})=a_{\bm{k}}^2P_n^{(1)}(\bm{k})$ from
\begin{align}
P_n(\bm{k})P_{+\uparrow}\ket{u_n^{(1)}(\bm{k})}&=a_{\bm{k}}P_n(\bm{k})\ket{\bm{k}=0,+,\uparrow}\notag\\
&=a_{\bm{k}}^2\ket{u_n^{(1)}(\bm{k})}.
\end{align}
Thus, we obtain
\begin{align}
P_n(\bm{k})P_{+\uparrow}P_n(\bm{k})&=P_n(\bm{k})P_{+\uparrow}[P_n^{(1)}(\bm{k})+P_n^{(2)}(\bm{k})]\notag\\
&=a_{\bm{k}}^2P_n^{(1)}(\bm{k}).
\end{align}
Taking the trace of the equality, we obtain
\begin{align}
a_{\bm{k}}^2&=\Tr[P_n(\bm{k})P_{+\uparrow}]=\Tr[P_n(\bm{k})P_+^\eta]/2.
\end{align}
The MCBB eigenstate is given by
\begin{align}
\ket{u_n^{(1)}(\bm{k})}&=\frac{1}{a_{\bm{k}}^2}P_n(\bm{k})P_{+\uparrow}\ket{u_n^{(1)}(\bm{k})}\notag\\
&=\frac{1}{a_{\bm{k}}}P_n(\bm{k})\ket{\bm{k}=0,+,\uparrow}.
\end{align}
$\ket{u_n^{(2)}(\bm{k})}$ is obtained by acting $\Theta I$.
We finally obtain
\begin{align}
\ket{u_n^{(\sigma)}(\bm{k})}&=\sqrt{\frac{2}{\Tr[P_n(\bm{k})P_+^\eta]}}P_n(\bm{k})\ket{\bm{k}=0,+,\sigma}.
\end{align}
Thus, the MCBB matrix elements are evaluated by the formula
\begin{align}
&\braket{u_n^{(\sigma)}(\bm{k})|A|u_{n'}^{(\sigma')}(\bm{k})}=\frac{2\Tr[P_+^\eta\ket{\sigma'}_s\bra{\sigma}_sP_n(\bm{k})AP_{n'}(\bm{k})]}{\sqrt{\Tr[P_n(\bm{k})P_+^\eta]\Tr[P_{n'}(\bm{k})P_+^\eta]}},\notag\\
&\ket{\sigma'}_s\bra{\sigma}_s=\delta_{\sigma\sigma'}\frac{1+\sigma s_z}{2}+\delta_{\sigma,-\sigma'}\frac{s_x-i\sigma s_y}{2}.
\end{align}

It is convenient to rewrite the matrix elements by using the Pauli matrices of the pseudospin,
\begin{align}
\!\!\!\!\braket{u_n^{(\sigma)}(\bm{k})|A|u_{n'}^{(\sigma')}(\bm{k})}&=[a^0_{nn'}(A)+\bm{a}_{nn'}(A)\cdot\bm{\sigma}]_{\sigma\sigma'}.\label{eq:def_matA}
\end{align}
The first term is given by
\begin{align}
a_{nn'}^0(A)&=\frac{1}{2}\sum_{\sigma}\braket{u_n^{(\sigma)}(\bm{k})|A|u_{n'}^{(\sigma)}(\bm{k})}\notag\\
&=\frac{\Tr[P_+^\eta P_n(\bm{k})AP_{n'}(\bm{k})]}{\sqrt{\Tr[P_n(\bm{k})P_+^\eta]\Tr[P_{n'}(\bm{k})P_+^\eta]}}.
\end{align}
The second term is given by
\begin{align}
\bm{a}_{nn'}(A)&=\frac{1}{2}\sum_{\sigma\sigma'}\bm{\sigma}_{\sigma'\sigma}\braket{u_n^{(\sigma)}(\bm{k})|A|u_{n'}^{(\sigma'}(\bm{k})}\notag\\
&=\frac{\Tr[P_+^\eta\bm{s}P_n(\bm{k})AP_{n'}(\bm{k})]}{\sqrt{\Tr[P_n(\bm{k})P_+^\eta]\Tr[P_{n'}(\bm{k})P_+^\eta]}}.
\end{align}
Thus, we arrive at the formula~\eqref{eq:formulas_me_A} in the main text.

The interband Berry connection in the MCBB gauge is given by $A\to \partial_{X_i}H_{\rm N}/(\epsilon_{n'}-\epsilon_n)$ in the formulas obtained above:
\begin{align}
[iA_{\inter}^{X_i}]_{nn'}&=\frac{(1-\delta_{nn'})\sigma_\mu}{\epsilon_{n'}(\bm{k})-\epsilon_n(\bm{k})}\label{eq:Ainter_formula}\\
&\qquad\cdot\frac{\Tr[P_+^\eta s^\mu P_n(\bm{k})\partial_{X_i}H_{\rm N}(\bm{k})P_{n'}(\bm{k})]}{\sqrt{\Tr[P_n(\bm{k})P_+^\eta]\Tr[P_{n'}(\bm{k})P_+^\eta]}}.\notag
\end{align}
The covariant derivative of the interband Berry connection $A^{X_i}_{\inter;X_j}$ can be obtained by using Eq.~\eqref{eq:Ainter_formula} and MCBB matrix elements of $A\to \partial_{X_i}\partial_{X_j}H_{\rm N}$ based on the formula Eq.~\eqref{eq:CD_formula}.
For example, for the case of $A^{q_i}_{\inter;q_j}$, which coincides with the covariant derivative in the wave-number space $A^i_{\inter;j}(\bm{k})$, we obtain
\begin{align}
&[iA^{i}_{\inter;j}]_{n,-n}(\epsilon_{-n}-\epsilon_n)\\
&=[w^{ij}]_{n,-n}-[iA^{i}_\inter,v_\intra^j]_{n,-n}-[iA^j_\inter,v_\intra^i]_{n,-n}.\notag
\end{align}
Here we defined
\begin{align}
[w^{ij}]_{n\sigma,n'\sigma'}&=\braket{u_n^{(\sigma)}|\partial_{k_i}\partial_{k_j}H_{\rm N}|u_{n'}^{(\sigma')}},\notag\\
[v^i_\intra]_{n\sigma,n'\sigma'}&=\delta_{nn'}\braket{u_n^{(\sigma)}|\partial_{k_i}H_{\rm N}|u_{n'}^{(\sigma')}}.
\end{align}
By using this, we obtain the covariant derivative of interband Berry connection in the spinful bilayer model as shown in Eq.~\eqref{eq:covariantderivative_spinfulbilayer}.

\section{Weakly noncentrosymmetric bilayer}
\label{app:weaklynoncentro}
The GBR is applicable to the cases where the two bands are entangled, i.e., degenerate at some discrete points in the Brillouin zone.
As an example, let us consider the weakly-noncentrosymmetric bilayer
\begin{align}
H_{\rm N}(\bm{k};V)&=\xi(\bm{k})+t_\perp(\bm{k})\eta_x+V\eta_z+\bm{g}(\bm{k})\cdot\bm{s}\,\eta_z,\label{eq:51}
\end{align}
by introducing the potential gradient $V$ to the Hamiltonian in Eq.~\eqref{eq:locallynoncentro}.
Here, $V$ is much smaller than the band gap without $V$, i.e., $R(\bm{k})=\sqrt{t_\perp(\bm{k})^2+g(\bm{k})^2}$, but can be larger than the applied Zeeman field.
We assume the even-parity order parameter
\begin{align}
\Delta(\bm{k})=\psi(\bm{k})+\bm{d}(\bm{k})\cdot\bm{s}\,\eta_z.\label{eq:52}
\end{align}
In the following, let us consider the response of this system to the Zeeman field $-\bm{h}\cdot\bm{s}$.

By using the results for GBR and QGPP in Appendix~\ref{app:QGPP_by_TRSP}, the normal-state part of the GBR$+$MCBB is given by
\begin{equation}
\bar{\epsilon}_n(V)=\epsilon_n+(\bm{g}_n-\bm{h}_n)\cdot\bm{\sigma}+O\left(h^2,[V/R]^2\right),
\end{equation}
after some calculations based on Eq.~\eqref{eq:energy_SM}.
This means that a spinful noncentrosymmetric system is effectively realized with the spin-orbit coupling
\begin{align}
\bm{g}_n&=\frac{nV}{R}\bm{g},
\end{align}
and the Zeeman field $\bm{h}_n$ given in Eq.~\eqref{eq:effective_Zeeman}.
We used the Berry connection for the potential gradient
\begin{align}
V[iA^V_{\inter}]_{n,-n}&=\frac{nt_\perp V}{2R^2}\hat{g}\cdot\bm{\sigma}.
\end{align}
QGPP is also obtained by, from Eqs.~\eqref{eq:QGPP_for_Y} and~\eqref{eq:Delta_g_general},
\begin{align}
\Delta_\geom(\bm{h},V)&=\delta_{nn'}\left[\frac{t_\perp V}{R^2}\left(\frac{nt_\perp}{R}\bm{d}_\parallel+\bm{d}_\perp\right)\cdot\bm{\sigma}\right]\\
&\quad+\delta_{n,-n'}\left[\frac{t_\perp g V}{R^3}\bm{d}_\parallel\cdot\bm{\sigma}\right]+[\Delta_{\rm e,\geom}(\bm{h})]_{nn'}\notag,
\end{align}
where we abbreviated $O(hV,h^2,V^2)$ terms for simplicity and $[\Delta_{\rm e,\geom}(\bm{h})]_{nn'}$ is given in Eq.~\eqref{eq:41}.
Thus, neglecting the components of the order parameter between $n$ and $-n$, the system is described by the assembly of the effective two-band models,
$\bar{H}_\B(\bm{h},V)=\oplus_{n=\pm}\bar{H}_{\B,n}$ with
\begin{align}
\!\!\!\!\!\bar{H}_{\B,n}\equiv\begin{pmatrix}
    \epsilon_n+(\bm{g}_n-\bm{h}_n)\cdot\bm{\sigma}&\psi_n+\bm{d}_n\cdot\bm{\sigma}\\
    \psi_n^*+\bm{d}_n^*\cdot\bm{\sigma}&-\epsilon_n-(\bm{g}_n+\bm{h}_n)\cdot\bm{\sigma}
\end{pmatrix},
\end{align}
and
\begin{subequations}
\begin{align}
\psi_n&=\psi+\frac{ng}{R}\bm{d}\cdot\hat{g}-i\frac{1}{R^2}\bm{h}\cdot\bm{d}\times\bm{g},\\
\bm{d}_n&=\frac{t_\perp V}{R^2}\left(\frac{nt_\perp}{R}\bm{d}_\parallel+\bm{d}_\perp\right).
\end{align}
\end{subequations}
In accordance with the inversion symmetry,
the effective order parameter is spin-singlet in the absence of $V$, while the spin-triplet component admixes in the presence of the potential gradient $V$.
By assuming $V\gg h$, we can further evaluate QGPP for Abelian cases as illustrated in Sec.~\ref{sec:Abelian}.
Based on the results for the two-band model, this system is topologically nontrivial when $\psi$ and $\bm{d}$ are the dominant $d$-wave pairing and $p$-wave pairing, respectively.
The above results are relevant to thin films of bilayer $d$-wave superconductors such as YBa$_2$Cu$_3$O$_7$ in proximity to ferromagnets and under the potential gradient introduced by e.g, gating techniques.

\end{document}